\documentclass[twocolumn,floatfix,altaffilletter,superscriptaddress,preprintnumbers,tightenlines,showpacs,showkeys,notitlepage,nofootinbib]{revtex4-1}
\pdfoutput=1

\usepackage[colorlinks=true,citecolor=blue,linkcolor=blue]{hyperref}
\usepackage[normalem]{ulem}
\usepackage{amsmath,amssymb}
\usepackage{mathtools}
\usepackage{verbatim}
\usepackage{epsfig}
\usepackage{graphicx}               
\usepackage{url}
\usepackage{color}
\usepackage{multirow}
\usepackage{placeins}
\usepackage[dvipsnames]{xcolor}
\usepackage{epstopdf}
\usepackage{siunitx}
\usepackage{fontawesome}

\usepackage{cleveref}
\usepackage{lipsum}
 \usepackage{gensymb}
 
\allowdisplaybreaks

\newcommand{\ev}[1]{\ensuremath{\left\langle #1 %
                     \right\rangle}} 

\newcommand{\BR}{\text{BR}}

\newcommand{\iso}[2]{{\ensuremath{{}^{#2}}\ensuremath{\rm #1}}}


\begin{document}

\title{The Neutrino Magnetic Moment Portal: \\ Cosmology, Astrophysics, and Direct Detection}

\author{Vedran Brdar}
\email{vbrdar@mpi-hd.mpg.de}
\affiliation{Max Planck-Institut f\"ur Kernphysik, Saupfercheckweg 1,
             69117 Heidelberg, Germany}

\author{Admir Greljo}
\email{admir.greljo@cern.ch}
\affiliation{Theoretical Physics Department, CERN,
             1 Esplanade des Particules, 1211 Geneva 23, Switzerland}

\author{Joachim Kopp}
\email{jkopp@cern.ch}
\affiliation{Theoretical Physics Department, CERN,
             1 Esplanade des Particules, 1211 Geneva 23, Switzerland}
\affiliation{PRISMA Cluster of Excellence \& Mainz Institute for
             Theoretical Physics, \\
             Johannes Gutenberg University, Staudingerweg 7, 55099
             Mainz, Germany}

\author{Toby Opferkuch}
\email{toby.opferkuch@cern.ch}
\affiliation{Theoretical Physics Department, CERN,
             1 Esplanade des Particules, 1211 Geneva 23, Switzerland}

\preprint{CERN-TH-2020-130,
          MITP/20-041}

\begin{abstract}
\noindent 
We revisit the physics of neutrino magnetic moments, focusing in particular on
the case where the right-handed, or sterile, neutrinos are heavier (up to
several MeV) than the left-handed Standard Model neutrinos.  The discussion is
centered around the idea of detecting an upscattering event mediated by a
transition magnetic moment in a neutrino or dark matter experiment. Considering
neutrinos from all known sources, as well as including all available data from
XENON1T and Borexino, we derive the strongest up-to-date exclusion limits on
the active-to-sterile neutrino transition magnetic moment.  We then study
complementary constraints from astrophysics and cosmology, performing, in
particular, a thorough analysis of BBN. We find that these data sets scrutinize
most of the relevant parameter space. Explaining the XENON1T excess with
transition magnetic moments is marginally possible if very conservative assumptions
are adopted regarding the supernova 1987\,A and CMB constraints. Finally, we
discuss model-building challenges that arise in scenarios that feature large
magnetic moments while keeping neutrino masses well below \SI{1}{eV}. We
present a successful ultraviolet-complete model of this type based on TeV-scale
leptoquarks, establishing links with muon magnetic moment, $B$ physics
anomalies, and collider searches at the LHC.
\end{abstract}

\maketitle

\section{Introduction}
\label{sec:intro}

When dark matter detection using nuclear recoils was first proposed by
Goodman and Witten in 1985 \cite{Goodman:1984dc}, the idea was presented
as a parasitical measurement in a neutrino detector proposed a few months
earlier \cite{Drukier:1983gj}. Since then, direct dark matter
searches have turned into a vast field of research in its own right,
with numerous advanced experiments and with a community numbering in
the thousands.  With the current generation of detectors, the circle
closes as dark matter experiments are becoming sensitive probes
of low-energy neutrino physics. In particular, experimental
sensitivities are approaching the so-called ``neutrino floor'', an
unavoidable background due to scattering of solar and atmospheric
neutrinos \cite{Gutlein:2010tq, Harnik:2012ni, Feng:2014uja}. The resulting
nuclear and electronic recoils are in general indistinguishable from a
dark matter signal, and are therefore often characterized as a severe
limitation to dark matter searches. On the other hand, precision
measurements of the neutrino floor
also offer tremendous discovery opportunities for phenomena beyond
the Standard Model (SM) in neutrino physics \cite{Harnik:2012ni}.

In this work, we focus in particular on neutrino magnetic moments,
which are predicted to be tiny ($< \SI{e-19}{\mu_B}$) in the
SM \cite{Fujikawa:1980yx, Lee:1977tib, Petcov:1976ff, Pal:1981rm, Shrock:1982sc,
Dvornikov:2003js, Giunti:2014ixa, Tanabashi:2018oca}, but can be substantially
larger in theories beyond the SM \cite{Shrock:1974nd,Shrock:1982sc, Fukugita:2003en,
  Giunti:2014ixa, Lindner:2017uvt, Babu:2020ivd,Xu:2019dxe}.  The possibility that neutrino magnetic
moments enhance the neutrino floor in direct dark matter detection experiments
has been discussed for quite some time \cite{Harnik:2012ni}, but only
now experiments are reaching the sensitivity to set meaningful
constraints \cite{Aprile:2020tmw}.

In view of these new experimental opportunities, our goals in this paper are
the following: in \cref{sec:mm}, we discuss the event spectra for solar
neutrinos recoiling
against electrons and nuclei in the presence of large neutrino magnetic
moments. Unlike much of the previous literature, we allow the mass of the
right-handed neutrinos to be much larger than the mass of their left-handed
partners, so that magnetic moment-induced scattering processes $\nu_L + e^- \to
N_R + e^-$ and $\nu_L + X_Z^A \to N_R + X_Z^A$ can be inelastic. Here, $N_R$
denotes the right-handed (sterile) neutrinos, and $X_Z^A$ is an atomic nucleus.
This well-motivated possibility opens up significant new parameter space.
We use our event
spectra to derive limits from XENON1T and Borexino
data~\cite{Aprile:2020tmw,Agostini:2018uly}, and to predict the sensitivity of future
observatories like DARWIN.  We also show that the excess electron recoil events
reported in ref.~\cite{Aprile:2020tmw} can be explained by neutrino transition
magnetic moments.  This possibility has also been considered in
refs.~\cite{Babu:2020ivd, Shoemaker:2020kji}, but compared to these papers, we
will employ a much more detailed fit, including a more sophisticated treatment
of backgrounds and covering a much larger recoil energy range. We find
qualitative differences compared to the results of ref.~\cite{Shoemaker:2020kji},
and we will discuss their origin.  It is worth mentioning
that after the announcement of the XENON1T excess in ref.~\cite{Aprile:2020tmw},
an avalanche of papers has appeared offering various explanations of the anomaly.
Without being exhaustive, let us mention a couple of promising scenarios, namely
dark photons~\cite{An:2020bxd, Bloch:2020uzh, Okada:2020evk}), inelastic dark
matter down-scattering~\cite{Bell:2020bes, Bramante:2020zos,Choudhury:2020xui} and light dark matter decay \cite{Farzan:2020dds}. A proposed
explanation in terms of solar axions is more difficult to realize, though see
\cite{DiLuzio:2020jjp, Gao:2020wer, Bloch:2020uzh}.

We will also go well beyond refs.~\cite{Babu:2020ivd, Shoemaker:2020kji} in
\cref{sec:constraints}, where we discuss a comprehensive set of constraints on
neutrino magnetic moments. In particular, we show how astrophysical
observations (stellar cooling, supernova 1987A), cosmological measurements
(BBN, CMB), and terrestrial experiments (neutrino scattering) disfavor vast
regions of parameter space, while nevertheless leaving large swathes open.  We
present in particular detailed simulations of Big Bang Nucleosynthesis (BBN)
in the presence of neutrino
magnetic moments.  We also outline model-building strategies for avoiding these
constraints, showing that some options are quite simple, while others are
fairly exotic.
In the final part of the paper, \cref{sec:lq}, we depart from the effective
field theory (EFT) description of neutrino magnetic moments and discuss
ultraviolet (UV) completions. Typically, large neutrino magnetic moments
require fine-tuning to avoid large corrections to the neutrino masses. We show
explicit models featuring TeV-scale leptoquarks, partially motivated by various
anomalies in $B$ physics, which elegantly avoid this problem.

We will summarize our results and conclude in \cref{sec:conclusions}.

\section{Neutrino Magnetic Moments and Direct Dark Matter Searches}
\label{sec:mm}

\subsection{Modified Solar Neutrino Spectrum}
\label{sec:solar-nu}

Neutrino magnetic moments are described at low energies by the effective operator,
\begin{align}
  \mathcal{L_\mu} = \frac{\mu_\nu^\alpha}{2} \, F_{\mu\nu} \, \bar\nu_L^\alpha \sigma^{\mu\nu} N_R 
                  + \text{h.c.}\,,
  \label{eq:L-mu}
\end{align}
where $F^{\mu\nu}$ is the electromagnetic field strength tensor, $N_R$ is a
right-handed neutrino gauge singlet, and $\nu_L^\alpha$ is the SM neutrino
field of flavor $\alpha$. In this paper, we will assume transition magnetic
moments between $\nu_\mu$ and $N_R$ unless otherwise stated. This is motivated
by the UV completions we consider in \cref{sec:lq}.  To simplify our notation,
we will omit the superscript $\alpha$ in the following.  The factor
$\tfrac{1}{2}$ in \cref{eq:L-mu} is a convention usually adopted in the
literature.  We imagine that the operator in \cref{eq:L-mu} originates from
short-distance new physics above the electroweak scale which generates the
following gauge-invariant operators
\begin{align}
  \mathcal{L}  \supset
      \frac{c_B}{\Lambda^2} g' B_{\mu\nu} \,\bar{L}_L \tilde{H} \sigma^{\mu\nu} N_R
    + \frac{c_W}{\Lambda^2} g W^a_{\mu\nu} \,\bar{L}_L \sigma^a \tilde{H} \sigma^{\mu\nu} N_R \,.
  \label{eq:L-muEFT}
\end{align}
Here $g$ ($g'$) is the $SU(2)_L$ (hypercharge) gauge coupling, $\Lambda$ is the cutoff scale,
$L_L$ denotes a SM lepton doublet, $W^a_{\mu\nu}$ ($B_{\mu\nu}$) is the
$SU(2)_L$ (hypercharge) field strength tensor, $\sigma^a$ are Pauli matrices, and $\tilde{H} \equiv i \sigma^2 H^*$
is the conjugate Higgs field.  After electroweak symmetry breaking, the neutrino
magnetic moment becomes
\begin{align}
  \mu_\nu = \frac{\sqrt{2} e \, v_H}{\Lambda^2} \, (c_B + c_W)\,,
  \label{eq:mu-nu}
\end{align}
with the Higgs vacuum expectation value $v_H$ and the electromagnetic gauge
coupling $e$.
The operator in \cref{eq:L-mu} mediates neutrino--electron scattering,
$\nu_L + e^-\to N_R + e^-$, as well as neutrino--nucleus scattering,
$\nu_L + X^A_Z \to N_R + X^A_Z$. Since the masses of $\nu_L$ and $N_R$ can
in general be different, the scattering may be inelastic.
The differential scattering rates for the two processes are \cite{Vogel:1989iv, Harnik:2012ni,Magill:2018jla,Shoemaker:2018vii,Balantekin:2013sda,Coloma:2017ppo}
\begin{widetext}
\begin{align}
  \frac{d\sigma_{\mu}(\nu_L e \to N_R e)}{dE_r}
    &= \alpha \mu_\nu^2 \bigg[
         \frac{1}{E_r} - \frac{1}{E_\nu}
       + M_N^2 \frac{E_r -2 E_\nu  - m_e}{4 E_\nu^2 E_r m_e}
       + M_N^4 \frac{E_r - m_e}{8 E_\nu^2 E_r^2 m_e^2}
       \bigg]
                                              \label{eq:sigma-mm-e} \\
\intertext{and}
  \frac{d\sigma_{\mu}(\nu_L X_Z^A \to N_R X_Z^A)}{dE_r}
    &= \alpha \mu_\nu^2 Z^2 F_1^2(E_r) \bigg[
         \frac{1}{E_r} - \frac{1}{E_\nu}
       + M_N^2 \frac{ E_r -2 E_\nu - m_X}{4 E_\nu^2 E_r m_X}
       + M_N^4 \frac{E_r - m_X}{8 E_\nu^2 E_r^2 m_X^2}
       \bigg] \nonumber\\
    &\quad
     + \alpha \mu_\nu^2 \mu_X^2 F_2^2(E_r) \bigg[
         \frac{2 m_X}{E_\nu^2} \Big( (2 E_\nu - E_r)^2 - 2 E_r m_X \Big)
       + M_N^2 \frac{E_r - 4 E_\nu}{E_\nu^2}
       + M_N^4 \frac{1}{E_\nu^2 E_r}
       \bigg] \,.
  \label{eq:sigma-mm-N}
\end{align}
\end{widetext}
Here, $M_N$ is the right-handed neutrino mass, $\alpha$ is the electromagnetic
fine structure constant, $E_\nu$ is the neutrino energy, $E_r$ is the electron
or nuclear recoil energy, and $Z$ is the nuclear charge in units of $e$. Also,
$m_X$ and $\mu_X$ are the nuclear mass and magnetic moment, respectively, while
$A$ is the number of nucleons. The term in the first line of \cref{eq:sigma-mm-N}
contains an enhancement
factor $Z^2$ because in low-energy scattering the neutrino
interacts with the whole nucleus coherently.  At higher energies, the
substructure of the nucleus is partly resolved and coherence is broken. This is
described by the nuclear charge and magnetic form factors $F_1(E_r)$,
$F_2(E_r)$.  The charge form factor can be parameterized as $F_1(E_r) = 3
e^{-\kappa^2 s^2/2} [\sin(\kappa r)-\kappa r\cos(\kappa r)] / (\kappa r)^3$,
with $s = \SI{1}{fm}$, $r = \sqrt{R^2 - 5 s^2}$, $R = 1.2 A^{1/3}$~fm, $\kappa
= \sqrt{2 m_X E_r}$ (and $q^2 \simeq -\kappa^2$)~\cite{Engel:1991wq}.  For the
magnetic form factor, no such general parameterization exists.  In the
following, we will \emph{neglect} nuclear magnetic moments (and thus the whole
second line of \cref{eq:sigma-mm-N}) because scattering on the nuclear magnetic
moment is strongly suppressed compared to scattering on the nuclear charge due
to the absence of $Z^2$ enhancement.

\Cref{eq:sigma-mm-e,eq:sigma-mm-N} should be compared to the corresponding
expressions for neutrino--electron scattering and neutrino--nucleus scattering
in the Standard Model \cite{Gutlein:2010tq, Harnik:2012ni, Lindner:2016wff,
Bednyakov:2018mjd}:
\begin{widetext}
\begin{align}
  \frac{d\sigma_\text{SM}(\nu_e e \to \nu_e e)}{dE_r} &=
  \frac{G_F^2 m_e}{2 \pi  E_\nu^2}
    \Big[ 4 s_w^4 (2 E_\nu^2+E_r^2-E_r (2 E_\nu+m_e))-2s_w^2 (E_r m_e-2 E_\nu^2)+
        E_\nu^2 \Big] \,,
  \label{eq:dsigmadE-nue-e} \\
  \frac{d\sigma_\text{SM}(\nu_{\mu,\tau} e \to \nu_{\mu,\tau} e)}{dE_r}  &=
  \frac{G_F^2 m_e}{2 \pi E_\nu^2}
    \Big[ 4 s_w^4 (2 E_\nu^2+E_r^2-E_r (2 E_\nu+m_e))+2s_w^2 (E_r m_e-2 E_\nu^2)+
        E_\nu^2 \Big] \,,
  \label{eq:dsigmadE-numu-e} \\
  \frac{d\sigma_\text{SM}(\nu_{e,\mu,\tau} X_Z^A \to \nu_{e,\mu,\tau} X_Z^A)}{dE_r} &=
  \frac{G_F^2 m_X Q_w^2 F^2(E_r)}{8 \pi E_\nu^2} \Big[
    2 E_\nu^2 - 2 E_\nu E_r - E_r m_X
  \Big] \,.
  \label{eq:dsigmadE-nu-N}
\end{align}
\end{widetext}
In the last expression, $Q_w = 2 T_3 - 4 Z \sin^2 \theta_w$ is the weak charge
of the nucleus, which depends on its weak isospin $T_3 = (2 Z - A) / 2$.  As
\cref{eq:dsigmadE-nu-N} accounts only for the vector couplings of the $Z$
boson, but not its axial-vector interactions, this expression is strictly valid
only for spin-0 nuclei.  We will, however, use it also for nuclei with non-zero
spin because vector interactions always dominate for heavy nuclei thanks to the
enhancement by $Q_w^2$.  Axial-vector couplings do not profit from such an
enhancement because the contributions from different nucleons tend to cancel,
rather than adding up coherently like for the vector couplings.

Note that in the literature, the term $2 E_\nu E_r$ in \cref{eq:dsigmadE-nu-N}
is often dropped because it is much smaller than the other two terms at the recoil
energies of experimental interest. In some references, an extra term $E_r^2$ is
included inside the square brackets. This term arises when the nucleus is
treated as a spin-$1/2$ fermion, but it should be omitted for
spin-0 nuclei.  In any case, this extra term is usually
negligibly small compared to the others.

The event rate is obtained from the differential cross section $d\sigma / 
dE_r$ according to~\cite{Harnik:2012ni}
\begin{align}
  \frac{dR}{dE_r} = N_T \epsilon(E_r) \int_{E_\nu^\text{min}} \! \textrm{d}E_\nu
                    \frac{\textrm{d} \Phi}{\textrm{d} E_\nu}
                    \frac{\textrm{d} {\sigma}}{\textrm{d} E_r}\,,
  \label{eq:theo-rate}
\end{align}
where $N_T$ is the number of target electrons, $\epsilon(E_r)$ is the
detection efficiency, and $\textrm{d} \Phi / \textrm{d} E_\nu$
is the neutrino flux. We adopt the solar neutrino flux from the BS05(OP)
Standard Solar Model (SSM)
introduced in ref.~\cite{Bahcall:2004pz}.  As we will mostly assume that
only one of the active neutrino flavors participates in the magnetic moment
interactions, we need to include also neutrino oscillations. We do so
following the standard approach (see for instance ref.~\cite{Akhmedov:1999uz}),
assuming fully adiabatic flavor transitions.
The lower integration  boundary in
\cref{eq:theo-rate} is the minimum neutrino energy required to attain a recoil
energy $E_r$,
\begin{align}
  E_\nu^\text{min}(E_r) = \tfrac{1}{2} \Big[ E_r + \sqrt{E_r^2 + 2 m_e E_r} \Big]
                          \bigg(1 + \frac{M_N^2}{2 E_r m_e} \bigg) \,.
  \label{eq:E-nu-min}
\end{align}

\begin{figure*}
  \begin{tabular}{cc}
    \includegraphics[width=0.49\textwidth]{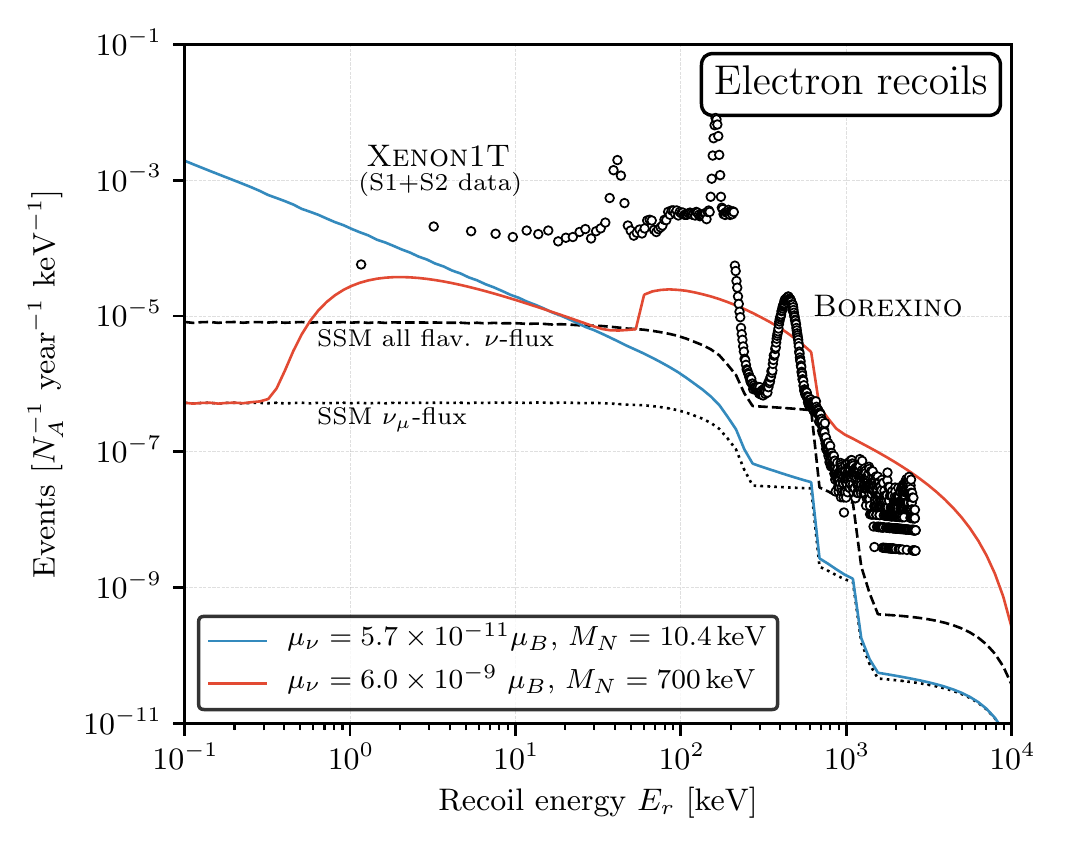} 
    \includegraphics[width=0.49\textwidth]{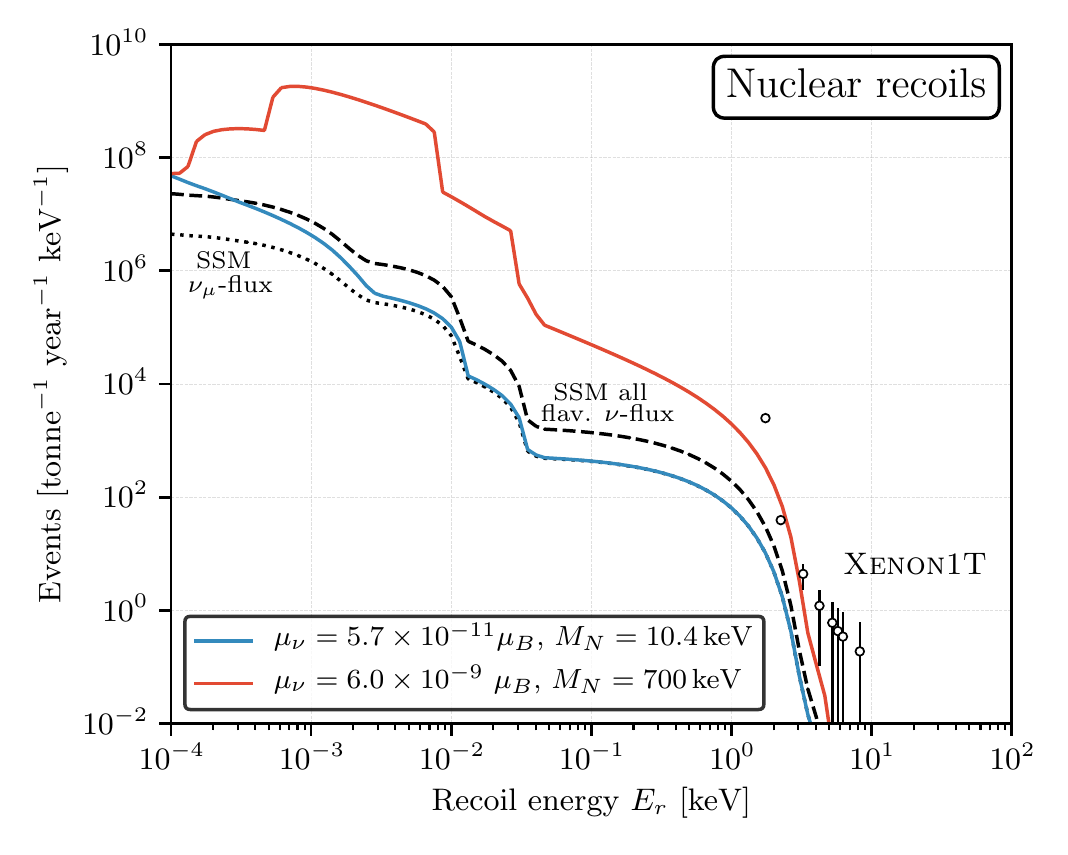}
  \end{tabular}
  \caption{The rate of neutrino-induced electron recoils (left) and
    nuclear recoils (right) in a dark matter detector for the Standard Model
    (black curves) and for scenarios with enhanced neutrino magnetic moments
    (colored curves).  We also include data points from XENON1T~\cite{Aprile:2020tmw}
    and Borexino~\cite{Agostini:2018uly} for comparison.  For very light
    right-handed neutrinos, we observe the $1/E_r$ scaling predicted by
    \cref{eq:sigma-mm-e,eq:sigma-mm-N} in that limit. For larger $M_N$,
    the predicted recoil spectrum has a broad bump, and the interplay of
    \cref{eq:sigma-mm-e,eq:sigma-mm-N} with the shape of the solar neutrino
    spectrum leads to interesting spectral features.
  }
  \label{fig:recoil-spectrum}
\end{figure*}

In \cref{fig:recoil-spectrum} we compare the electron and nuclear recoil rates
in the SM to those predicted in the presence of non-zero neutrino magnetic
moments, for two combinations of $(M_N, \mu_\nu)$.  For the case of
massless or very light $N_R$ (blue curve), the scaling with $1/E_r$ at low
recoil energies predicted by \cref{eq:sigma-mm-e,eq:sigma-mm-N} is evident.
The chosen value of $\mu_\nu = \SI{5.7e-11}{\mu_B}$ corresponds
to the best fit point to the XENON1T low-energy excess (see \cref{sec:xenon} below).
It is important, however, to emphasize,
that the stellar cooling limits discussed below in \cref{sec:stellar-cooling}
disfavor such large $\mu_\nu$ by more than an order of
magnitude~\cite{Arceo-Diaz:2015pva} for $M_N \lesssim \SI{20}{keV}$.

If $M_N$ is larger, comparable to the
center-of-mass energy $\sqrt{s} = \sqrt{m_e^2 + 2 E_\nu m_e}$, the magnetic
moment-induced recoil spectrum changes shape: it vanishes below
\begin{align}
  E_r^\text{min} = \frac{M_N^2}{2( m_e + M_N)}
  \label{eq:Er-min}
\end{align}
because no kinematic solutions exist for any
$E_\nu$.  At higher $E_r$, the spectrum gradually increases before
reaching a maximum, which for a monochromatic neutrino flux is located at
a recoil energy of
\begin{align}
  E_r^\text{peak}(E_\nu)
    = \frac{2 m_e M_N^4}{8 E_\nu^2 m_e^2 - 2 m_e M_N^2 (2 E_\nu + m_e) + M_N^4} \,.
  \label{eq:Er-peak}
\end{align}
The maximum possible recoil energy for a given neutrino energy $E_\nu$
is reached at
\begin{multline}
  E_r^\text{max}(E_\nu)
    = \frac{1}{2 E_\nu + m_e}
      \bigg[
        E_\nu^2
      - \tfrac{1}{2} M_N^2 \\
      + \frac{E_\nu}{2 m_e} \Big(
            \sqrt{M_N^4 - 4 M_N^2 m_e (E_\nu + m_e) + 4 E_\nu^2 m_e^2} - M_N^2 \Big)
      \bigg] \,.
  \label{eq:Er-max}
\end{multline}
As different components of the solar neutrino spectrum switch on at different
values of $E_r$ (see \cref{eq:E-nu-min}), the event spectrum strongly depends
on $M_N$ whenever $M_N^2 / (2 E_r m_e) \gtrsim 1$. In
\cref{fig:recoil-spectrum}, we illustrate this by showing the spectrum for one
representative parameter point with large $M_N$ (red curve).  For the chosen
case, the XENON1T excess is accommodated by the \iso{Be}{7} neutrino flux, and the
higher-energy CNO, pep, \iso{B}{8}, and hep fluxes lead to a
substantial number of recoil events also at higher energies. While these
higher-energy events are also visible in XENON1T, they are probed with
much greater sensitivity by Borexino, as can be seen by comparing to the XENON1T
and Borexino data points which we have included in \cref{fig:recoil-spectrum}
for comparison.  This example illustrates the more general statement that
XENON1T will be most sensitive at low $M_N$, while Borexino is in a better
position to constrain models with $M_N \gtrsim \mathcal{O}(\SI{100}{keV})$.

\subsection{XENON1T}
\label{sec:xenon}

To compare the predicted magnetic moment signals in XENON1T to the data more
quantitatively, we have carried out maximum-likelihood fit to both
electron recoil and nuclear recoil data.  Electron recoils in
XENON1T have recently attracted significant attention due to a
$\sim 3\sigma$ excess over background \cite{Aprile:2020tmw}, while nuclear recoils
are the main channel for direct detection of Weakly Interacting Massive Particle (WIMP)
dark matter, and the channel for which dark matter detectors are optimized
\cite{Aprile:2018dbl}.

We use binned electron recoil event rates in the complete energy range from zero to
\SI{210}{keV}. For the low energy range we use the data from fig.~4 of
ref.~\cite{Aprile:2020tmw}, while for the range from \SI{30}{} to \SI{210}{keV}
we take the data from fig.~3 of the same reference. The background predictions
for the individual components are again lifted from fig.~3 of ref.~\cite{Aprile:2020tmw},
but in the fit we allow their normalization to float within the uncertainties listed
in Table~I of ref.~\cite{Aprile:2020tmw}.  The neutrino signal is composed
of the SM weak scattering processes described by
\cref{eq:dsigmadE-nue-e,eq:dsigmadE-numu-e}, and of the new physics piece
given by \cref{eq:sigma-mm-e}.

The event rate follows \cref{eq:theo-rate}, with the detection efficiencies
taken from Fig.~2 of ref.~\cite{Aprile:2020tmw}.  To obtain $N_T$ for the case
of XENON1T, we sum over all stable isotopes of xenon, weighted by their natural
abundances. After computing the expected count rates according to
\cref{eq:theo-rate}, we apply Gaussian smearing based on the detector
resolution given in fig.~2 of the supplemental material to
ref.~\cite{XENON:2019dti}.

For predicting the rate of nuclear recoils in XENON1T, we construct
bins in $E_r$ from the S1 (scintillation) and S2 (ionization) signals,
see fig.~3 in ref.~\cite{Aprile:2018dbl}. The data of interest lies in the nuclear
recoil signal region in this figure. We employ Poissonian smearing in S1. The
number of photoelectrons in the S2 channels is much larger than in the S1 channel,
so the relative importance of Poisson fluctuations in that channel is much smaller,
so we neglect it.  
The most relevant background process arises from neutrons \cite{Aprile:2013tov},
which we include in our likelihood analysis. Detection efficiencies are taken
from fig.~1 of ref.~\cite{Aprile:2018dbl}.

\begin{figure*}
  \includegraphics{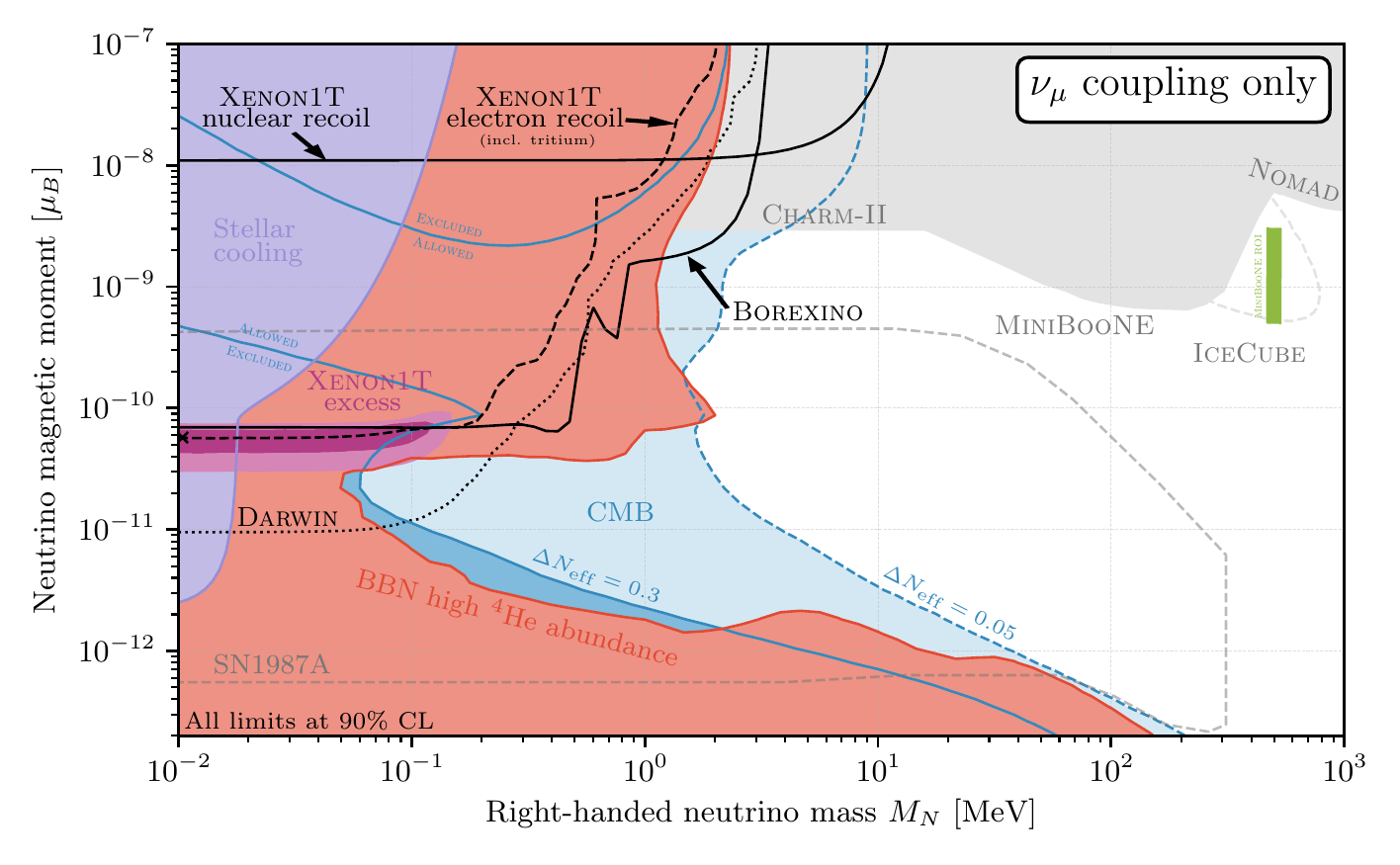}
  \caption{Allowed 90\% CL regions of the $\nu_\mu$--$N$ transition magnetic moment
    as a function of the right-handed neutrino mass $M_N$. We compare the XENON1T
    exclusion contour including a possible tritium contamination (black dashed1)
    to the limits we obtain from Borexino data (black solid) and to the
    projected sensitivity of DARWIN
    \cite{Aalbers:2016jon}. We also show in pink the $1\sigma$ and $2\sigma$
    regions preferred by the XENON1T excess in the absence of tritium contamination
    \cite{Aprile:2020tmw}.  We see that the excess region
    is consistent with constraints from stellar cooling
    (purple region, see \cref{sec:stellar-cooling}) and BBN (red shaded region,
    see \cref{sec:bbn}), as well as a conservative CMB constraint on $N_\text{eff}$
    (blue shaded region, see \cref{sec:cmb}). The motivation for considering this
    more conservative constraint is the fact that larger $N_\text{eff}$ is preferred
    by local measurements of the Hubble constant $H_0$. A less conservative
    CMB constraint (light blue) is in tension with the XENON1T-preferred region.
    The constraint from SN1987A (dashed gray contours) could be in tension with
    XENON1T as well, but we emphasize that this constraint may not be robust
    (see ref.~\cite{Bar:2019ifz}, reviewed in \cref{sec:sn1987A}).  We finally
    include terrestrial constraints (gray region at the top of the plot), taken
    from ref.~\cite{Magill:2018jla,Coloma:2017ppo} as well as sensitivity projections for Icecube \cite{Coloma:2017ppo} (dashed) and the region preferred by MiniBooNE data (green) \cite{Magill:2018jla}.
  }
  \label{fig:constraints}
\end{figure*}

The results of our fit to XENON1T data are shown in \cref{fig:constraints}.
At low $M_N$, our fit to electron recoil data is consistent with the one carried
out by the XENON collaboration in ref.~\cite{Aprile:2020tmw}, but note that
we have here assumed that only $\nu_\mu$ participate in the magnetic moment
interaction, whereas the XENON1T collaboration has assumed flavor-universal couplings.
XENON1T Constraints on transition magnetic moments involving $\nu_e$ and $\nu_\tau$
are similar to the ones for $\nu_\mu$.
We find that magnetic moments $\mu_\nu \gtrsim \SI{7e-11}{\mu_B}$
are disfavored, and that the observed excess is well explained for $\mu_\nu$ slightly
below that value. Going beyond the analysis in ref.~\cite{Aprile:2020tmw}, we find
that the sensitivity deteriorates at $M_N \gtrsim \SI{200}{keV}$ as the lowest energy
solar neutrinos no longer have sufficient energy to create $N_R$.

Comparing our results to ref.~\cite{Shoemaker:2020kji}, we find a slightly
larger best fit value for $\mu_\nu$ at low $M_N$, which can again be understood
from our assumption that only $\nu_\mu$ feel the magnetic moment interaction.
At higher $M_N$, the authors of ref.~\cite{Shoemaker:2020kji} find secondary
best fit regions, the most prominent of which lies at $M_N \simeq \SI{600}{keV}$
and $\mu_\nu \simeq \SI{2e-9}{\mu_B}$.
Interactions in this secondary region would be dominated by \iso{B}{8} neutrinos.
This region does not appear in our results due to enhancements of the spectrum
at recoil energies $> \SI{30}{keV}$, see for instance the red curve
in \cref{fig:recoil-spectrum}. The plateau at $\sim \SI{100}{keV}$ that
disfavors the $M_N \simeq \SI{600}{keV}$ solution in our fit is driven
by the pep neutrino flux.

We have also entertained the likely possibility that the XENON1T excess is not a sign
of new physics, but of some yet unknown SM background.  Following the XENON1T
collaboration \cite{Aprile:2020tmw}, we use tritium decays as a proxy for such
a background, Including this background in our fit, we find the 90\%~CL
limit shown in \cref{fig:constraints} as a dashed black line.
In the future, we expect this limit to improve by about a factor 4 with the DARWIN
experiment~\cite{Aalbers:2016jon}. We have computed the sensitivity of
DARWIN assuming the same background model as in XENON1T (including tritium) 
and an exposure of $\SI{200}{t \cdot years}$. The result is shown as a
black dotted curve in \cref{fig:constraints}.

$N_R$ masses up to several tens of MeV are in principle accessible using nuclear
recoils in direct detection experiments, but the reach in $\mu_\nu$ is much
poorer in this case. The resulting limit is shown as a black curve
in \cref{fig:constraints} and has been cross-checked against a limit
derived using Bayesian statistics in ref.~\cite{Shoemaker:2018vii}.

Comparing to astrophysical and cosmological constraints, we find
that the parameter region most interesting to direct detection experiments
is only partially probed by stellar cooling constraints. At $M_N \gtrsim \SI{20}{keV}$,
these constraints become unimportant as production of such heavy $N_R$ is
suppressed even in the hot cores of red giant stars.
XENON1T's preferred region is in possible tension with the
anomalous cooling constraint from supernova 1987A and with the CMB measurement
of $N_\text{eff}$ (the effective number of relativistic degrees of freedom at
recombination).  We will discuss these constraints in more detail in
\cref{sec:constraints}, but we mention already here why these limits do not
necessarily spoil the explanation of the XENON1T excess in terms of neutrino transition
magnetic moments. For SN1987A, the argument is based on ref.~\cite{Bar:2019ifz},
which argues that the observed neutrino signal from SN1987A might not have  been
from the cooling core, but rather from matter being accreted onto that core. Such a neutrino
flux would not be significantly altered in the presence of $N_R$.  For $N_\text{eff}$,
early Universe observables (Planck, Baryon Acoustic Oscillations) considered in
isolation prefer a value very close to 3 (dashed blue contour in
\cref{fig:constraints})~\cite{Philcox:2020vvt}.
Considering, however, late-time measurements of the Hubble constant $H_0$,
somewhat larger values are preferred as they help relax the tension between
late-time and early-time measurements of $H_0$~\cite{Bernal:2016gxb, Knox:2019rjx}.
Another very strong constraint is imposed by BBN (red shaded region in 
\cref{fig:constraints}, see \cref{sec:bbn} for details). This constraint is
only marginally consistent with the XENON1T excess.

\Cref{fig:constraints} also summarizes terrestrial constraints, especially from
CHARM-II, MiniBooNE, and NOMAD~\cite{Shoemaker:2018vii} (see \cref{sec:terrestrial} for
details). These limits are most relevant for $M_N \gtrsim
\si{MeV}$.  In this mass range, they strongly constrain the large-$\mu_\nu$
parameter region.  We
emphasize that the limits from CHARM-II, MiniBooNE, and NOMAD are relevant only for
transition magnetic moments between $\nu_\mu$ and $N_R$. For $\nu_e$--$N_R$ and
$\nu_\tau$--$N_R$ couplings, these experiments do not set a constraint. In \Cref{fig:constraints} we also include Icecube sensitivity \cite{Coloma:2017ppo} as well as the region preferred by 
MiniBooNE data \cite{Magill:2018jla}.

\subsection{Borexino}
\label{sec:borexino}

XENON1T's constraints on neutrino magnetic moments are complemented by those
based on neutrino--electron scattering in Borexino.
We have recast the analysis performed by Borexino in ref.~\cite{Agostini:2018uly}
for the low-energy recoil region. Analogously to XENON1T the expected differential
event rate is given by \cref{eq:theo-rate}. The key differences are the increased
number of target electrons $N_T = \num{3E+31}$ and the detector efficiency, which
for $E_r \in [200,2600]~\si{keV}$ is $\epsilon(E_r) \sim 1$.
To determine the (Gaussian) energy resolution, we use the calibration data
shown in fig.~21 of ref.~\cite{Agostini:2017aaa}, based on injecting mono-chromatic
photons from the sources given in tab.~II of the same reference. Borexino provides
the full data underpinning fig.~2 of ref.~\cite{Agostini:2018uly} as well as a
mapping between the number of PMT hits $N_h$ and the recoil energy of the event
in digital form~\cite{BxOpen}.  We find a $5\%$ mismatch between the given
$N_h$--$E_r$ mapping and the one obtained through fitting the calibration
data in fig.~21 of ref.~\cite{Agostini:2017aaa} in the region where the
\iso{Be}{7} and $pep$ neutrino fluxes dominate. In our analysis, we correct for
this mismatch by correcting energies 5\% upwards. We find that, with this
corrections, our predictions for the solar neutrino event rates in the SM
match Borexino's predictions very accurately.

To determine the 90\%~CL exclusion limit on the neutrino magnetic moment we
perform a binned likelihood fit to the data from
refs.~\cite{Agostini:2018uly,BxOpen}, similar to the fit to XENON1T data
described in \cref{sec:xenon} above.  For Borexino we allow the individual
background components as well as the normalization of the various solar
neutrino fluxes to vary. The best-fit point is very well consistent with
$\mu_\nu = 0$.  The exclusion contour shown in \cref{fig:constraints} is
obtained by assuming $\Delta\chi^2$ to follow a $\chi^2$ distribution with two
degrees of freedom. As a cross-check we compare our results to the constraint
on a flavor-universal magnetic moment obtained in ref.~\cite{Borexino:2017fbd}.
We find $\mu_\nu < \SI{3.1E-11}{\mu_B}$ at 90\%~CL in this case, compared to
$\SI{2.8E-11}{\mu_B}$ in ref.~\cite{Borexino:2017fbd}.  Note that the
constraint in \cref{fig:constraints} is for couplings to muon neutrinos only,
therefore the limit appearing in the figure is weaker.

There are two key differences between the electron recoil constraints from
Borexino and XENON1T. First, at low $M_N$, XENON1T is more sensitive due its
lower recoil energy threshold, which allows it to more efficiently probe
the $1/E_r$-enhanced flux at low energies.  Secondly, Borexino
has greater sensitivity at larger $M_N$ values due to the larger recoil
energies measured in the detector, resulting in smaller $E_\nu^\text{min}$
cut-off values (see \cref{eq:E-nu-min}).  Of course, also the larger size of
Borexino plays an important role.

\section{Constraints}
\label{sec:constraints}

\subsection{Stellar Cooling}
\label{sec:stellar-cooling}

\subsubsection{Magnetic Moment Constraints from Stellar Cooling}

Inside the hot plasma forming the core of a star, the dispersion relation of
electromagnetic excitations (called plasmons) is modified in such
a way that two-body decays into $\nu_L + N_R$, induced by the magnetic
moment operator from \cref{eq:L-mu}, can become kinematically allowed.  For
this to happen, it is required that $N_R$ is lighter than the core
temperature of the star.  As $\nu_L$ and $N_R$ can escape the star
unhindered, they would constitute an efficient energy sink, greatly increasing
the rate at which the star loses energy.  Such excess energy loss, if too
large, would be grossly inconsistent with our understanding of stellar
evolution. Even under very conservative assumptions, tight limits on neutrino
magnetic moments can thus be derived.  For instance, to maintain its observed
temperature in spite of the extra energy loss, the star would need to burn more
fuel and would thus exhaust its fuel supply sooner.  This way, limits on
$\mu_\nu$ can be derived from the simple fact that the Sun is still around
after burning for 4.5~billion years. Stronger limits can be derived from a more
detailed comparison to stellar evolution models \cite{Raffelt:1996wa,
Raffelt:1999gv}.  As the core temperature of the Sun is of order keV, these
limits extend up to right-handed neutrino masses of the same order.  The reach
in $M_N$ can be extended by about an order of magnitude by looking at the
evolution of red giant stars \cite{Raffelt:1994ry, Raffelt:1996wa,
Arceo-Diaz:2015pva, Diaz:2019kim}: at the tip of the red giant branch in the
Hertzsprung--Russell diagram, where the helium flash occurs, the core temperature
is on the order of \SI{10}{keV}.

Starting from \cref{eq:L-mu}, we compute the decay rate of transverse plasmons
into $\nu + N_R$ in the comoving frame,
\begin{align}
  \Gamma_{\gamma^*} = \frac{|\mu_\nu|^2 K^4}{24 \pi \, \omega}
                      \bigg(1 - \frac{M_N^2}{K^2} \bigg)^2
                      \bigg(1 + 2\, \frac{M_N^2}{K^2} \bigg) \, \theta(K-M_N) \,,
\end{align}
where $\theta(x)$ is the Heaviside step function, $\omega$ and $k$ are the plasmon
energy and momentum, respectively, and $K = \sqrt{\omega^2 - k^2}$ is the effective
plasmon mass. The energy loss per unit volume is (see Appendix~B of
ref.~\cite{Vogel:2013raa})
\begin{align}
  Q = \int_0^\infty \! \frac{k^2 dk}{\pi^2} \int_{M_N^2}^\infty \! \frac{d \omega^2}{\pi} \,
      \frac{\omega \, \Gamma_T}{(K^2-\omega_p^2)^2 +  (\omega \Gamma_T)^2} \,
      \frac{\omega \, \Gamma_{\gamma^*}}{e^{\omega/T_\gamma}-1} \,,
  \label{eq:loss}
\end{align}
where $\Gamma_T = 8 \pi \alpha^2 n_e / 3 m_e^2$ is the Thomson scattering rate.
The plasma characteristics for a red-giant core just before helium ignition are
taken from Table~D.1 of ref.~\cite{Raffelt:1996wa}. In particular, $\omega_p = \SI{18}{keV}$,
$T_\gamma = \SI{8.6}{keV}$ and $n_e = \SI{3e29}{cm^{-3}}$. We rely on the
recent analysis of globular clusters in ref.~\cite{Diaz:2019kim}, which sets
an upper limit on the active neutrino magnetic moment $|\mu_\nu | < \SI{2.2e-12}{\mu_B}$.
In order to recast this bound for the case of massive $M_N$, we equate the energy
loss in \cref{eq:loss} for massive and massless $N$, and solve for the unknown
$\mu_\nu (M_N)$. The obtained limits on $\mu_\nu$ as a function of $M_N$ are shown
in fig.~\ref{fig:constraints} as the purple exclusion region. For $M_N$ smaller than
the plasma frequency $\omega_p$, the dominant effect comes from on-shell plasmon
decays and the propagator in \cref{eq:loss} can effectively be approximated by
the delta function $\propto \delta(K^2 - \omega_p^2)$. Instead, when $M_N > \omega_p$,
the cooling process $\gamma + e^- \to e^-  + N + \nu$ quickly becomes
phase-space suppressed with increasing $M_N$.

\subsubsection{Avoiding Stellar Cooling Constraints}
\label{sec:stellar-cooling-2}

Even though stellar cooling bounds are extremely robust, they are evaded in
models in which the properties of the $N_R$ depend dynamically on the
surrounding matter density.  Such ``chameleon'' models have been proposed
originally to explain dark energy \cite{Khoury:2003rn}, but have also been
invoked in other contexts, for instance to avoid stellar cooling
constraints on axion-like particles \cite{Masso:2005ym, Jaeckel:2006xm,
Ganguly:2006ki, Kim:2007wj, Brax:2007ak, Redondo:2008tq, Bloch:2020uzh}.
Consider a very light scalar field $\phi$ coupled to $N_R$ through an
operator of the form
\begin{align}
  \mathcal{L}_{\phi R} \supset \lambda_{\phi R} \phi \, \overline{N_R^c} N_R \,,
  \label{eq:L-chameleon}
\end{align}
with a dimensionless coupling constant $\lambda_\phi$.
This operator implies that the $N_R$ mass will be larger in environments
of large $N_R$ density such as the cores of stars.  At the edge of the solar
core (radius $0.1 R_\odot$), the $N_R$ mass shift due to the operator in
\cref{eq:L-chameleon} will be of order
\begin{align}
  \Delta m_{RR} &\sim \SI{0.8}{keV} \times \lambda_{\phi R}^2
                                    \bigg( \frac{\si{meV}}{m_\phi} \bigg)^2
                                    \bigg( \frac{n_R}{\SI{e11}{cm^{-3}}} \bigg)
  \label{eq:mass-shift-chameleon-R}
\end{align}
for an $N_R$ that would otherwise saturate the solar cooling bound.
To arrive at this estimate, we have assumed that all emitted $N_R$ have an energy
of order $T_\odot \sim \SI{1.5e6}{Kelvin}$, that the total $N_R$ luminosity
equals the solar luminosity, and that the coupling $\lambda_\phi$ is of order one.

If $\phi$ couples  not only to $N_R$, but also to SM quarks, the mass shift
becomes proportional to the number density of SM fermions, which in the solar
core ($\rho \simeq \SI{150}{grams/cm^3}$) is 15 orders
of magnitude larger than the $N_R$ number density that went into
\cref{eq:mass-shift-chameleon-R}, $n_{N_R} \simeq \SI{e11}{cm^{-3}}$.
On the other hand, $\phi$ couplings to SM fermions would typically proceed through
mixing with the Higgs boson, which would lead to extra suppression by the small
fermion Yukawa couplings and by the Higgs mixing angle $\theta$, which is typically of
order the mass ratio $m_\phi / m_H$.  Writing the $\phi$ coupling to quarks as
$\mathcal{L}_{\phi q} \supset \sum_q \lambda_{\phi q} (\sqrt{2} m_q / v_H) (m_\phi / m_H) \, \phi
\, \bar{q} q$, an additional contribution to the $N_R$ mass of order
\begin{align}
  \Delta m_{Rq} &\sim \SI{0.008}{keV} \times \lambda_{\phi R} \lambda_{\phi q}
                                      \bigg( \frac{\si{meV}}{m_\phi} \bigg)
                                      \bigg( \frac{m_q}{\SI{93}{MeV}} \bigg) \, 
  \label{eq:mass-shift-chameleon-q}
\end{align}
ensues.
To arrive at \cref{eq:mass-shift-chameleon-q}, we have used the density of the solar core,
$\rho_\odot = \SI{150}{grams/cm^3}$~\cite{Bahcall:2004pz}, the SM Higgs mass
$m_H = \SI{125}{GeV}$, and the SM Higgs vev $v_H = \SI{246}{GeV}$ \cite{Tanabashi:2018oca}.
For the quark content of the nucleon, $\sum_q m_q \ev{\bar{q} q}$, we have used
the numbers from ref.~\cite{Bali:2016lvx}. We see that the mass shift due
to the coupling to quarks will dominate for $m_\phi \gtrsim \si{eV}$ and for
small $n_R \lesssim \SI{1e8}{cm^{-3}}$. In these parameter regions, however, the magnitude
of the shift is too small to evade stellar cooling constraints.  For that, only
the small-$m_\phi$ regime ($m_\phi \ll \si{eV}$) is interesting, where the
contribution from \cref{eq:mass-shift-chameleon-q} can be safely neglected compared
to the one from \cref{eq:mass-shift-chameleon-R}.

\subsection{Supernova 1987A}
\label{sec:sn1987A}

The reasoning that leads to stellar cooling constraints also applies to
supernovae.  The production of $N_R$ through a magnetic moment operator would
open up an efficient energy sink, which in turn would lead to faster cooling of
the proto-neutron star~\cite{Raffelt:1996wa, Magill:2018jla}.  As a
consequence, thermal emission of neutrinos would decline faster, and the
duration of the observed neutrino burst would be shorter. Given that neutrinos
from supernova 1987A were observed for about 10~seconds \cite{Hirata:1987hu,
Alekseev:1988gp, Bionta:1987qt}, bounds can be derived on the neutrino magnetic
moment and on the mass of the right-handed neutrinos. The corresponding
exclusion region (dashed gray contour in \cref{fig:constraints}), taken from
ref.~\cite{Magill:2018jla}, shows several characteristic features: obviously,
very low $\mu_\nu \lesssim \text{few} \times \SI{e-13}{\mu_B}$ cannot be
constrained because the rate of $N_R$ production is too small to be detectable
in this case.  However, the region with $\mu_\nu \gtrsim \text{few} \times
\SI{e-10}{\mu_B}$ cannot be constrained either. There, $N_R$ interact too
frequently to leave the supernova core, in spite of being copiously produced.
Therefore, they do not contribute efficiently to the cooling of the core.
Finally, the constraint peters out at $M_N \gtrsim \SI{100}{MeV}$, where $N_R$
are too heavy to be produced in plasmon decays.

While supernova constraints are part of the standard canon in studies
of neutrino magnetic moments (and other manifestations of new physics
at scales $\lesssim \SI{100}{MeV}$), they have recently been called
into question \cite{Bar:2019ifz}.  The argument is that there is no
evidence the neutrinos observed from SN~1987A were
actually emitted from the core region of the exploding star. Rather,
a hot accretion disk may have formed around a fast-rotating core,
and the observed neutrino flux may have originated from this disk.  As an
accretion disk (unlike a supernova core) is optically thin not only to
$N_R$, but also to $\nu_L$, the production of $N_R$ through a magnetic
moment interaction would not lead to significant extra energy loss.

Even if the concerns raised in ref.~\cite{Bar:2019ifz} should be
disproven with future observations, supernova constraints can still
be relaxed by dedicated model-building along the lines of
\cref{sec:stellar-cooling-2}.  In particular, chameleon-like $N_R$,
whose mass and/or couplings differ in the dense environment of a
supernova core from those in vacuum can be invoked to prevent the
production of $N_R$ in supernovae, or to trap them inside to avoid
excess cooling.

\subsection{Cosmology}
\label{sec:cosmo}

Just as in stars and supernovae, the magnetic moment operator from \cref{eq:L-mu}
also opens a channel for producing $N_R$ in the early Universe. For  values
of $\mu_\nu$ large enough to be observable, $N_R$ will always thermalize
in the early Universe, unless the reheating temperature is extremely low.
The main cosmological consequences of the resulting population of $N_R$
are twofold:
\begin{itemize}
  \item While $N_R$ are relativistic, they contribute to the expansion rate,
    often parameterized in terms of the effective number of neutrino species,
    $N_\text{eff}$.
  \item $N_R \to \nu_L + \gamma$ decays inject extra photons into the Universe.
\end{itemize}
$N_\text{eff}$ is measured both at recombination and at the BBN epoch,
with the CMB allowing a deviation of about $0.3$ from the SM value
$N_\text{eff} = 3.045$ at 95\%~C.L.~\cite{Aghanim:2018eyx, deSalas:2016ztq,Akita:2020szl}\footnote{For $N_\text{eff}$ evaluation in SM and beyond, see \cite{Escudero:2018mvt,Escudero:2020dfa}.}
and BBN being even more restrictive ($N_\text{eff} \lesssim 3.2$ at 95\% C.L.)
\cite{Cyburt:2015mya}.  This implies that $N_R$ should either never thermalize,
or they should decay away before BBN,
or they should decouple above the electroweak scale so that subsequent
entropy production in the SM sector dilutes them sufficiently to satisfy
these constraints. A naive estimate for the $N_R$ decoupling temperature
$T_\text{dec}$
is obtained by equating the $N_R$ production rate $\sim \alpha \mu_\nu^2 T^3$
to the Hubble rate $\sim T^2 / M_\text{Pl}$, where $M_\text{Pl}$ is the
Planck mass. This yields
\begin{align}
  T_\text{dec} \simeq \SI{1.28}{GeV} \times
                      \bigg( \frac{\SI{e-11}{\mu_B}}{\mu_\nu} \bigg)^2 \,.
  \label{eq:T-dec}
\end{align}
This shows that sufficiently early decoupling is only achieved if
$\mu_\nu$ is two to three orders of magnitude below the sensitivity
of XENON1T.

From the $N_R$ lifetime,
\begin{align}
  \hspace{-0.5cm}
  \tau_N &= \frac{16 \pi}{\mu_\nu^2 M_N^3} \notag\\
         &= \SI{3\,760}{sec} \times \bigg( \frac{\SI{1e-11}{\mu_B}}{\mu_\nu} \bigg)^2 
                                    \bigg( \frac{\si{MeV}}{M_N} \bigg)^3 \,,
  \label{eq:Gamma-R}
\end{align}
we see that \si{MeV}-scale $N_R$ will typically \emph{not} decay before BBN.
They will, however, decay before recombination if
$\mu_\nu$ is in the observable range. Therefore, the CMB is sensitive to
$\mu_\nu$ only through $N_\text{eff}$, while the impact of a neutrino magnetic
moment on BBN is more involved~\cite{Cadamuro:2011fd, Depta:2020wmr}.
We will now describe in detail how we derive BBN limits on neutrino transition
magnetic moments.

\subsubsection{Big Bang Nucleosynthesis}
\label{sec:bbn}

If $N_R$ decays occur before the formation of heavy elements
($T \simeq \SI{100}{keV}$), but after the freeze-out of weak interactions that
can convert protons to neutrons ($T \simeq \SI{1}{MeV}$),
their main effect is through the modified expansion rate. Their presence in the
Universe means that $p \leftrightarrow n$ interactions freeze out faster,
leading to a larger neutron-to-proton ratio. Moreover, their decays alter the
expansion and cooling rates and thus the time available for neutrons to decay
and for BBN to proceed.  If $N_R$ decays happen after BBN, the extra photons
from $N_R$ decay decrease the baryon-to-photon ratio $\eta$.  As $\eta$ is
precisely measured during recombination, this effect needs to be compensated by
a larger $\eta$ during BBN. Larger $\eta$ renders deuterium disintegration less
efficient.  Once again, the presence of $N_R$ before the onset of BBN implies
that $p \leftrightarrow n$ reactions freeze out faster, and that neutrons have
less time to decay. All three effects imply larger abundances of heavy
elements.

To make these statements more quantitative, we have used a modified version
\cite{Depta:2020wmr} of the AlterBBN code \cite{Arbey:2011nf, Arbey:2018zfh}
that was kindly provided to us by Paul Frederik Depta, Marco Hufnagel, and
Kai Schmidt-Hoberg, who developed it to constrain axion-like particles in
ref.~\cite{Depta:2020wmr}. As input, this code requires a table listing
the relation between cosmological time $t$, the photon temperature $T_\gamma$,
the neutrino temperature $T_\nu$, the Hubble parameter $H$, and the
photon number density.  We obtain these quantities by solving the
following set of (integrated) Boltzmann equations that describe the evolution
of the photon, electron+positron, SM neutrino, and $N_R$ energy
densities $\rho_\gamma$, $\rho_e$, $\rho_\nu$, and $\rho_N$:
\begin{align}
  \begin{split}
    \dot\rho_\gamma &= - 4 H \rho_\gamma
                       + \ev{\sigma v}_{ee} (n_e \rho_e - n_e^\text{eq} \rho_e^\text{eq})
                       + \tfrac{1}{2} \Gamma_N (\rho_N - \rho_N^\text{eq}) \,, \\
    \dot\rho_e      &= - s_e H \rho_e
                       - \ev{\sigma v}_{ee} (n_e \rho_e - n_e^\text{eq} \rho_e^\text{eq}) \,, \\
    \dot\rho_\nu    &= - 4 H \rho_\nu
                       + \tfrac{1}{2} \Gamma_N (\rho_N - \rho_N^\text{eq}) 
                       + \Gamma_{eN} (\rho_N - \rho_N^\text{eq}) \,, \\
    \dot\rho_N      &= - s_N
                       - \Gamma_N (\rho_N - \rho_N^\text{eq})
                       - \Gamma_{eN} (\rho_N - \rho_N^\text{eq})\,.
  \end{split}
  \label{eq:bbn-boltzmann}
\end{align}
We now discuss the terms in these equations one by one.
The first term in each equation describes dilution and redshifting due to
Hubble expansion.  For the massive species, this term includes a factor
$s(m, T)$, which accounts for the transition from a relativistic
to a non-relativistic species. It is given by
\begin{align}
  s_i \equiv \frac{T}{\int \! dp \, f_i(m_i, T_i, p)}
  \int \! dp \, \frac{\text{d}f_i(m_i, T_i, p)}{\text{d}T} \,,
  \label{eq:s-factor}
\end{align}
where $f_i(m_i, T_i, T_{i0}, p)$ is the phase space distribution function of
species $i = e, N_R$. For electrons and positrons this is just the equilibrium
distribution. For $N_R$, we use the equilibrium distribution while the rates
of $N_R \leftrightarrow \gamma+\nu$ and $N_R + e \leftrightarrow \nu_L + e$
are larger than the Hubble rate. Otherwise, we use the distribution function
for a species that has decoupled while still relativistic.  As we will see that
there are no parameter points at which $N_R$ decouples while non-relativistic,
this approximation is sufficient for our purposes.

In the terms describing $e^+ e^-$ annihilation in \cref{eq:bbn-boltzmann}, we
approximate the annihilation cross section as $\ev{\sigma v}_{ee} \approx
\min[\alpha^2 / T_\gamma^2, \alpha^2 T_\gamma^2 / (4 m_e^4)]$, with $\alpha$ the electromagnetic
fine structure constant and $T_\gamma$ the photon temperature. This very rough
approximation is sufficient to ensure that the $e^+ e^-$ energy density follows
its equilibrium value,
\begin{align}
  \rho_e^\text{eq} = (2 T_\gamma^4 / \pi^2) J_f(m_e / T_\gamma) \,,
  \label{eq:rho-e-eq}
\end{align}
where $J_f(x)$ is a normalized integral over the Fermi-Dirac distribution:
\begin{align}
  J_f(x) \equiv \int \! dy \, \frac{y^2 \sqrt{x^2 + y^2}}
                                   {\exp(\sqrt{x^2 + y^2}) + 1} \,.
  \label{eq:Jf}
\end{align}
The equilibrium number density $n_e$ of electrons and positrons is defined
in an analogous way as
\begin{align}
  n_e^\text{eq} = (2 T_\gamma^3 / \pi^2) I_f(m_e / T_\gamma) \,,
  \label{eq:n-e-eq}
\end{align}
with
\begin{align}
  I_f(x) \equiv \int \! dy \, \frac{y^2}{\exp(\sqrt{x^2 + y^2}) + 1} \,.
  \label{eq:If}
\end{align}
We calculate the actual number density, $n_e$ from the energy density
according to $n_e = \rho_e / T_\gamma \cdot I_f(m_e/T_\gamma) / J_f(m_e/T_\gamma)$.

There are two terms in \cref{eq:bbn-boltzmann} that can alter the number density
of $N_R$. The one containing $\Gamma_R = 1 / \tau_R$ describes $N_R \to \gamma \nu$
decays and their inverse.  This term contains the $N_R$ equilibrium density $\rho_R^\text{eq}$,
which we compute in analogy to \cref{eq:rho-e-eq}, with the obvious replacement
$m_e \to M_N$ and the perhaps not so obvious replacement $T_\gamma \to
T_R = \frac{1}{2} (T_\nu + T_\gamma)$. Choosing this value for the $N_R$ equilibrium
temperature, we account for the fact that $N_R$ decays to both neutrinos and
photons.  There is also a term describing the $2 \leftrightarrow 2$ process
$e + N_R \leftrightarrow e + \nu_L$.  We obtain the corresponding rate $\Gamma_{eN}$
by computing the thermally averaged cross section $\ev{\sigma v}_{eN}$
and multiplying by $n_e$.

The treatment of the cosmological evolution outlined here is of course simplified
-- a more precise calculation would track not the integrated energy densities
$\rho_{\gamma,e,\nu,R}$, but rather the individual phase space distribution functions.
We have, however, verified that we can reproduce quite well the results of
ref.~\cite{Depta:2020wmr} for the case of axion-like particles (ALPs), especially
for the \iso{He}{4} abundance. The latter is measured by the parameter $\mathcal{Y}_p \equiv
\rho(\iso{He}{4}) / \rho_b$ which gives the \iso{He}{4} mass fraction relative
to the total baryonic mass density $\rho_b$. We have also verified that we
correctly predict $N_\text{eff} = 3.0$ in the SM at the CMB epoch.  (The small
correction to $N_\text{eff}$ stemming from $e^+ e^-$ annihilation into
neutrinos is ignored here as we will never use the absolute value of
$N_\text{eff}$, but only differences with respect to the SM.)

\begin{figure*}
  \centering
  \begin{tabular}{c@{\quad}c}
    \includegraphics[width=0.48\textwidth]{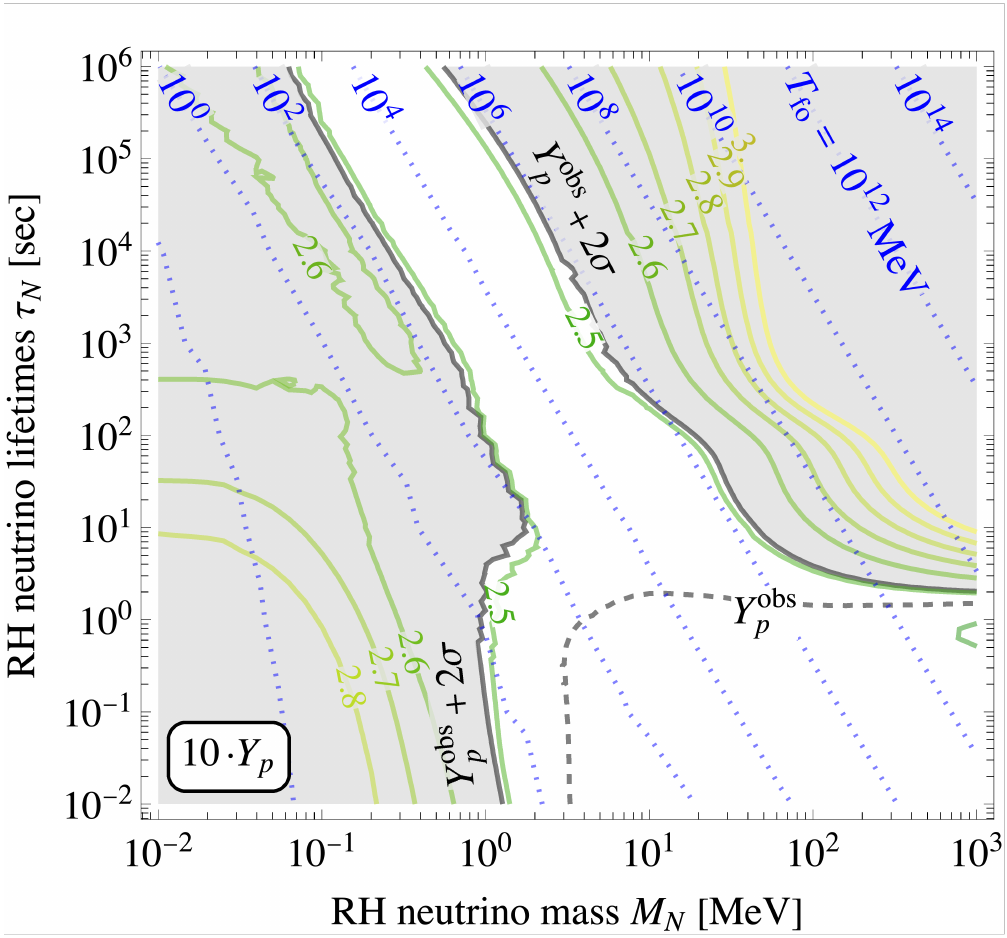} &
    \includegraphics[width=0.48\textwidth]{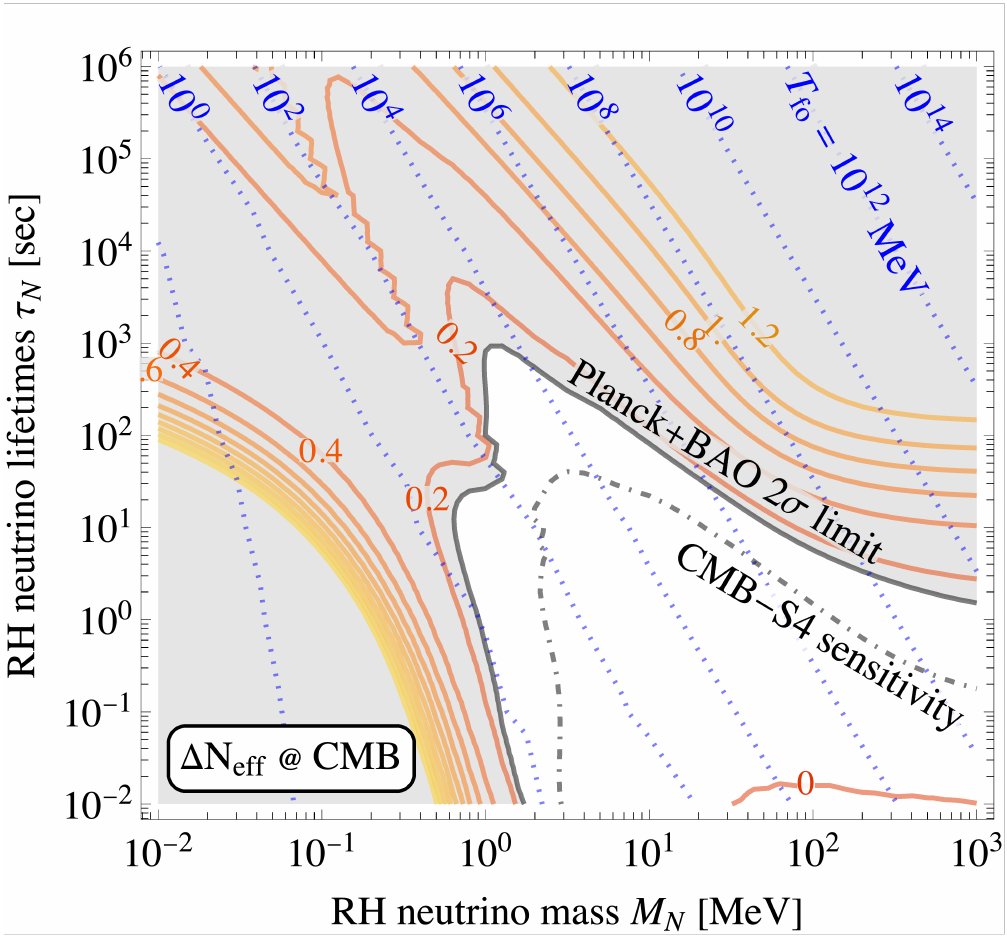} \\[0.2cm]
    (a) & (b)    
  \end{tabular}
  \caption{(a) The predicted \iso{He}{4} mass fraction $\mathcal{Y}_p \equiv
    \rho(\iso{He}{4}) / \rho_b$ in the presence of right-handed neutrinos
    of mass $M_N$, decaying after a lifetime $\tau_N$ via
    $N_R \to \nu_L + \gamma$. The observational constraint $\mathcal{Y}_p = 0.245
    \pm 0.006$ ($2\sigma$) is taken from ref.~\cite{Tanabashi:2018oca}.
    We also show, as blue dotted contours, the
    temperature at which $N_R$ freeze out from the thermal plasma (before recoupling
    later via $N_R \leftrightarrow \nu_L + \gamma$).  The unsteady behavior in the
    upper left-hand part of the plot is due to numerical instability.
    (b) The predicted deviation of $N_\text{eff}$, the effective number of relativistic
    degrees of freedom at recombination, from the SM expectation.
    We use $N_\text{eff} = 2.90^{+0.30}_{-0.32}$ ($2\sigma$) for the
    current limit~\cite{Philcox:2020vvt}, and $N_\text{eff} = 3.045 \pm 0.06$ 
    ($2\sigma$) for the expected sensitivity of CMB-S4~\cite{Abazajian:2016yjj}.
    The blue contours in this panel are the same as in panel (a).}
  \label{fig:Yp}
\end{figure*}

We plot our predictions for $\mathcal{Y}_p$ as a function of the right-handed
neutrino mass and lifetime in \cref{fig:Yp}~(a). The corresponding constraint
is also shown in \cref{fig:constraints} as a red shaded region.
Several different regimes are
apparent, which can be understood given the arguments at the
beginning of this section. At large mass and long lifetime, $N_R$ decouple
early, but remain abundant long after they have turned non-relativistic. The
Universe thus enters a fully or partially matter-dominated phase, where it
expands faster. Thus, $p \leftrightarrow n$ interactions freeze out faster and
there is thus less time for neutrons to decay, implying larger abundances of
the heavy elements.  Moreover, when the $N_R$ eventually decay, a large amount
of energy is deposited in the photon and neutrino baths, decreasing the
baryon-to-photon ratio $\eta$ at recombination. To compensate, $\eta$ must have
been larger during BBN, rendering deuterium disintegration less efficient and
once again contributing to larger abundances of the heavier elements.  In the
opposite limit of sub-MeV right-handed neutrinos, the $N_R$ always decay after
BBN.  Once again, their presence during BBN implies a larger expansion rate,
and their decays alter the baryon-to-photon ratio, leading to
more efficient production of heavy elements.  It is interesting that there is a
sweet spot for MeV-scale $N_R$ decoupling at a temperature of around
\SI{10}{GeV}. In this case, the $N_R$ decouple sufficiently early to be
depleted by entropy production during the QCD phase transition, while at the
same time their mass is low enough to never dominate the energy density of the
Universe.

It is important to note, though, that this sweet spot exists only for
$\mathcal{Y}_p$.  It is disfavored by the $N_\text{eff}$ measurement at the
CMB epoch (see \cref{fig:Yp}~(b) and discussion below in \cref{sec:cmb}), and
also by the measured deuterium abundance.  We do not include the latter in our
suite of constraints because our predictions for this quantity are slightly
less accurate than the ones for $\mathcal{Y}_p$ (as determined by comparing the
  results from our code to the ones from ref.~\cite{Depta:2020wmr} for the case
of ALPs).  We attribute this to the fact that we only track energy densities
rather than full distribution functions.

We also do not consider the $\iso{He}{3}$ abundance, which
would only lead to a very weak constraint in the upper right-hand corner of the
$M_N$--$\tau_N$ plane. The $\iso{He}{3}$ abundance can be measured only
locally, so its primordial abundance, which must be smaller than the local one,
remains unknown.

\subsubsection{Cosmic Microwave Background}
\label{sec:cmb}

Besides BBN, also the Cosmic Microwave Background (CMB) constrains the
existence of $N_R$, in particular through the measurement of the effective
number of degrees of relativistic degrees of freedom, $N_\text{eff}$.  Our
solution to \cref{eq:bbn-boltzmann} also yields a prediction for $N_\text{eff}$
at the time of recombination. As we have made simplifications in our treatment
of $e^+ e^-$ annihilation (in particular by neglecting $Z$-mediated
annihilation to neutrinos), we consider only the difference $\Delta
N_\text{eff}$ between the value of $N_\text{eff}$ predicted by solving
\cref{eq:bbn-boltzmann} in the presence of non-zero $\mu_\nu$ and the value
$N_\text{eff} = 3.0$ predicted by these same equations in the SM.  Our
predictions for $\Delta N_\text{eff}$ are shown in \cref{fig:Yp}~(b), and the
corresponding constraints are also included in \cref{fig:constraints} as blue
contours.

The qualitative behavior observed in \cref{fig:Yp}~(b) is similar to the one
we have seen in \cref{fig:Yp}~(a).  At large $M_N$ and long lifetime, the
Universe enters a phase of matter domination. The $N_R$ eventually decay after
neutrino decoupling, injecting half of the decay energy into the photon bath
and the other half into neutrinos. As by that time the energy density of
neutrinos is smaller by a factor $(4/11)^{4/3} \simeq 0.26$ than the energy
density of photons, the relative impact onto the neutrinos is larger, thus
$N_\text{eff}$ increases.  At low $M_N$ and small $\tau_N$, a rather
interesting phenomenon occurs: As both the decay $N_R \leftrightarrow \gamma +
\nu$ and the inelastic scattering process $N_R + e \leftrightarrow \nu + e$
remain in equilibrium for a very long time, they keep photons and neutrinos in
equilibrium for much longer than in the SM. $N_\text{eff}$ thus overshoots the
SM value by a large amount.  Between these two extremes, there is again a sweet
spot where $N_R$ decouple at about $T_\text{dec} \simeq \SI{10}{GeV}$,
sufficiently early to be substantially depleted during the QCD phase
transition. At the same time, their mass is too low to compensate this
depletion. Therefore, in this parameter region, the energy density of $N_R$
never plays an important role during the evolution of the early Universe.

Besides setting a limit on $N_\text{eff}$, CMB observations also constrain
neutrino magnetic moments in a more direct way. In particular, the CMB
spectrum is sensitive to any extra energy injected at late times by $N_R$
decays~\cite{Slatyer:2016qyl, Poulin:2016anj}. However, fig.~5 in
ref.~\cite{Poulin:2016anj} shows that the corresponding limits are always
weaker than BBN constraints for $N_R$ lifetimes $< \SI{e12}{sec}$. As
lifetimes longer than this are outside the range of our plots, we will
ignore these limits in the rest of this paper.

\subsubsection{Avoiding Cosmological Constraints}

Avoiding or weakening BBN and CMB constraints is not easy from a model-building point
of view.  One strategy is to prevent right-handed neutrino production in the
early Universe altogether, for instance by coupling $N_R$ to a
scalar field $\varphi$ whose vev evolves slowly over cosmological
history \cite{Bezrukov:2017ike, Farzan:2019yvo}.  The corresponding operator
$\lambda_\varphi \varphi \, \overline{N_R^c} N_R$ generates a
dynamical contribution to the $N_R$ mass. If $M_N \gg T$ in the early
Universe, $N_R$ production is forbidden and cosmological constraints
are avoided.

Another possibility is the introduction of a second, invisible, decay mode for
$N_R$ which dominates over $N_R \to \nu_L + \gamma$. That way,
injection of extra photons is avoided and only constraints on $N_\text{eff}$
are relevant.  As discussed above, these are satisfied if $N_R$ decoupling
is pushed to sufficiently early times, or if extra entropy is produced
between $N_R$ decoupling and BBN.  One possibility for such an extra
decay mode could be $N_R \to N_R' + \phi$, where $N_R'$ is a
second SM-singlet fermion and $\phi$ is a singlet scalar. Both $N_R'$
and $\phi$ would need to be significantly lighter than $M_N$, and the
coupling $\lambda_\phi \phi \bar{N}_R N_R'$ should be large. More precisely,
the decay rate for $N_R \to N_R' + \phi$ is
\begin{multline}
  \Gamma(N_R \to N_R' + \phi)
    = \frac{\lambda_\phi^2}{32 \pi M_N^3} (M_N^2 + M_N'^2 - m_\phi^2) \\
      \times \sqrt{(M_N^2 - M_N'^2 + m_\phi^2) - 4 M_N^2 m_\phi^2} \,.
  \label{eq:Gamma-R-phi}
\end{multline}
In the limit $M_N',\; m_\phi \to 0$, this becomes
\begin{align}
  \Gamma(N_R \to N_R' + \phi)
    \simeq \SI{6.6e-20}{sec} \times \lambda_\phi^2 \Big( \frac{M_N}{\si{MeV}} \Big) \,.
  \label{eq:Gamma-R-phi-approx}
\end{align}
That is, even for fairly small $\lambda_\phi$, rapid decays can be
realized.  While a detailed study of such a scenario is beyond the scope
of this work, qualitatively we see that if $M_N \gtrsim \si{MeV}$, the $N_R$
can be made to decay away before BBN becomes sensitive to their presence.
For smaller $M_N$, at least the injection of extra photons can be avoided.

\subsection{Terrestrial Constraints}
\label{sec:terrestrial}

While astrophysical and cosmological probes are particularly sensitive to
relatively small neutrino magnetic moments, terrestrial experiments are
essential for constraining parameter regions with large $M_N$ and/or large
$\mu_\nu$, where astrophysics and cosmology are often not sensitive. For large
$M_N$, this is because the temperatures at which the constraining processes
occur are too low to produce $N_R$.  At large $\mu_\nu$, stellar cooling
arguments are ineffective even for small $M_N$ because $N_R$, while copiously
produced in stars, would be trapped and therefore would not contribute
significantly to stellar energy loss.

On Earth, $N_R$ can be produced in beam dump experiments via upscattering of
light neutrinos in the detector or close to it ($\nu_L + X_Z^A \to N_R + X_Z^A$),
followed by the decay $N_R \to \nu_L + \gamma$ inside the fiducial volume.
Moreover, $N_R$ can be produced in meson decays close to the beam dump.
In collider experiments, the main production channels are
$e^+ e^- \to N_R \nu_L$ (LEP), $\bar{q} q \to N_R \nu_L$, and
$\bar{q} q' \to N_R \ell$ (LHC), where $q$, $q'$ are SM quarks
and $\ell$ is a SM lepton.

The terrestrial constraints shown in \cref{fig:constraints} are taken
from the compilation in ref.~\cite{Shoemaker:2018vii}.  They are based on
refs.~\cite{Geiregat:1989sz, Coloma:2017ppo} for CHARM-II,
on refs.~\cite{Magill:2018jla,AguilarArevalo:2007it} for MiniBooNE,
and on refs.~\cite{Altegoer:1997gv, Coloma:2017ppo} for NOMAD.

\section{Ultra-Violet Completion}
\label{sec:lq}

In many ultraviolet (UV) extensions of the SM, transition magnetic moments
between active and sterile neutrinos within the reach of direct dark matter
detection experiments are disfavored or require severe fine-tuning. The reason
is that large magnetic moments are often accompanied by prohibitively large
contributions to the active neutrino masses. More precisely, when removing the
photon line from the loop diagrams generating the magnetic moment, one
typically obtains a contribution to the neutrino mass renormalization. In the
following, we discuss the degree of fine-tuning required to circumvent this
problem, and we show how tuning can be avoided in models with TeV-scale
leptoquarks.

\begin{figure}
  \begin{tabular}{c@{}c}
    \includegraphics[width=0.9\columnwidth]{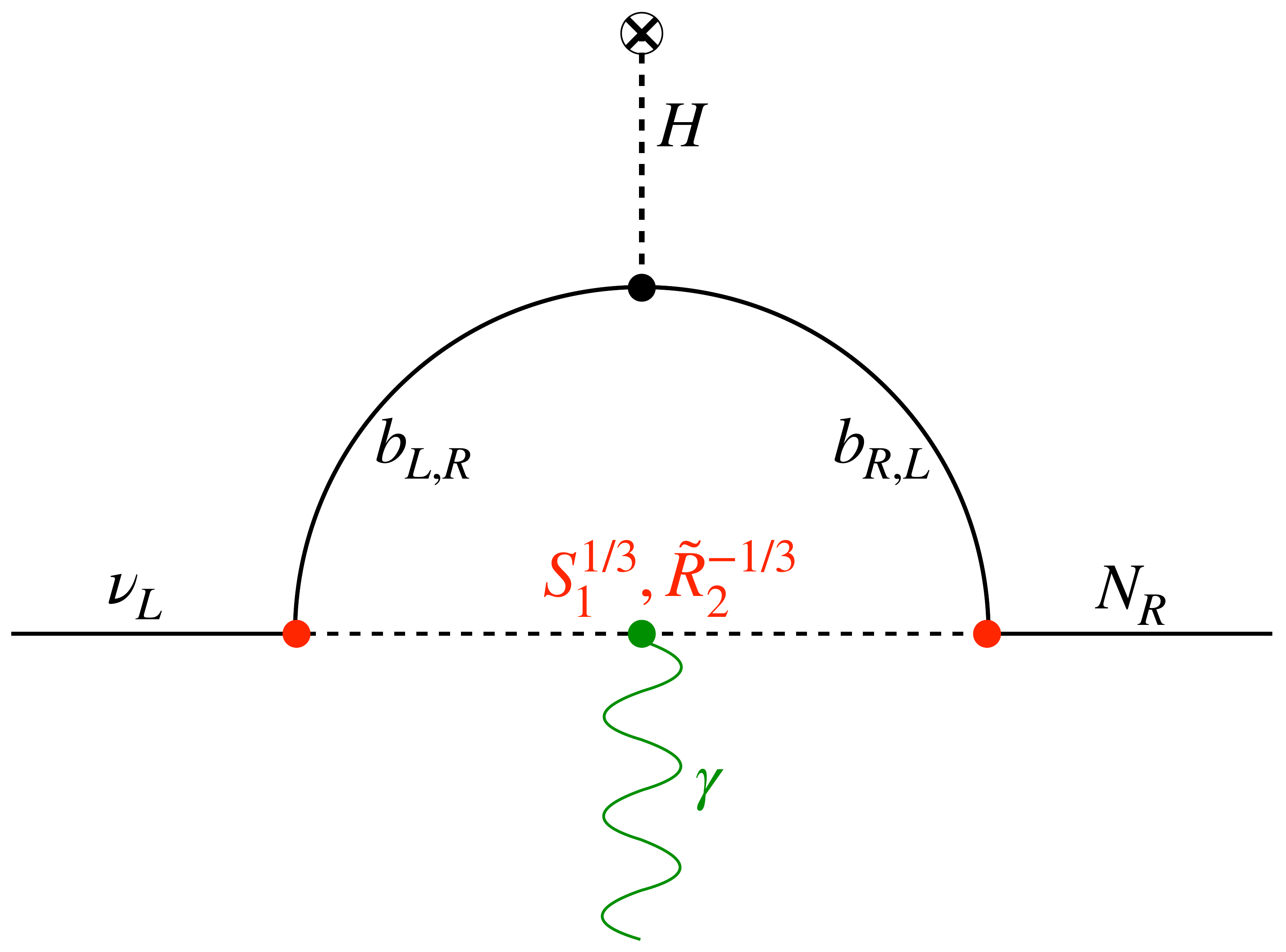} \\
  \end{tabular}
  \caption{The one-loop contribution to the neutrino magnetic moment induced by
    \cref{eq:L-lq-1,eq:L-lq-2}.  Note that both $S_1^{1/3}$ and $\tilde{R}_2^{-1/3}$
    can run in the loop. If the photon line is removed, these diagrams
    also contribute to neutrino masses. Combining $S_1^{1/3}$ and $\tilde{R}_2^{-1/3}$
    into a doublet under a horizontal $SU(2)_H$ symmetry, their one-loop 
    contributions to neutrino masses cancel exactly, while no cancellation occurs
  for the magnetic moment.}
  \label{fig:diagram}
\end{figure}

\textbf{Neutrino masses and magnetic moments} ---  Given that the loop diagrams
generating active-to-sterile transition magnetic moments $\mu_\nu$ and the ones
inducing Dirac masses $m_{\nu N}$ are often closely related (see for instance
\cref{fig:diagram}), a natural relation between $\mu_\nu$ and $m_{\nu N}$ is given by
\begin{align}
  \frac{\mu_\nu}{\mu_B} \approx \frac{m_e \,m_{\nu N}}{\Lambda^2} \,,
  \label{eq:mu-mass-relation}
\end{align}
where $\Lambda$ is the mass scale of the UV completion.
To recall, a Dirac mass term has the form  $\mathcal{L} \supset m_{\nu N} \,
\bar\nu_L N_R$, and the magnetic moment operator is $\mathcal{L} \supset
\tfrac{1}{2} \mu_\nu \, \bar \nu_L \sigma_{\mu \nu} N_R \, F^{\mu \nu}$, see \cref{eq:L-mu}.
For new physics at the electroweak scale, and $\mu_\nu$ within reach of
XENON1T, the natural expectation is $m_{\nu N} \gtrsim \mathcal{O}(\si{MeV})$.
Only by allowing for tuning between $m_{\nu N}$ and other contributions to the
active neutrino masses, the desired neutrino mass range $m_\nu \ll \SI{1}{eV}$
can be achieved.  Additional contributions to neutrino masses can arise, for
instance, from the type-I seesaw
mechanism~\cite{Minkowski:1977sc,Mohapatra:1979ia,Yanagida:1979as,GellMann:1980vs}.
In this case, the generic expectation for the RH neutrino masses is $M_N
\approx m_{\nu N}^2 / m_\nu$, far above the mass range of interest to dark
matter experiments.  Tuning the tree-level Yukawa coupling $\mathcal{L}_Y
\supset y_\nu \, \bar L_L \tilde{H} N_R$ such that the resulting tree-level
contribution to the Dirac mass, $y_\nu v_H/\sqrt{2}$, nearly cancels the
loop-induced contribution, the RH neutrino masses can be lowered to the MeV
scale while keeping the active neutrino masses $m_\nu \ll \si{eV}$. As a side
comment, note that the magnetic moment interaction also contributes to the
Majorana mass term of $\nu_L$ at one-loop, see Fig.~3 in
ref.~\cite{Magill:2018jla}. This contribution is of order $\mu_\nu^2 \Lambda^2
\, M_N / (16 \pi^2)$, so for the values of $\mu_\nu$ and $M_N$ that are of
interest to XENON1T and Borexino, and for $\Lambda \sim $~TeV, it is negligible.

In the inverse seesaw mechanism~\cite{Mohapatra:1986aw, Mohapatra:1986bd,
GonzalezGarcia:1988rw}, lepton number conservation makes active neutrinos
(almost) massless, while the sterile neutrino is a massive (pseudo) Dirac
fermion. Thus, two gauge singlet Weyl fermions, $N_L$ and $N_R$, are present.
Gauge symmetry and lepton number conservation allow the mass terms $\mathcal{L}
\supset - m_{\nu N} \bar \nu_L N_R - m_N \bar N_L N_R \, + \, \text{h.c.}$. The
mixing $\tan \theta_{\nu N} = m_{\nu N} / m_N$ rotates gauge to mass
eigenstates, predicting one massless chiral fermion and one massive Dirac
fermion with mass $\sqrt{m_{\nu N}^2 + m_{N}^2}$. To obtain the observed
non-zero masses for the active neutrinos, a small violation of lepton number is
needed on top of this. Thus, the inverse seesaw mechanism can bring the sterile
neutrino mass down to the range interesting for dark matter experiments without
tuning, but at the expense of introducing a large active--sterile mixing angle,
$\theta_{\nu N}$. The laboratory constraints on $\theta_{\nu N}$ from weak
interactions depend on the sterile neutrino mass and the active flavor it mixes
with. The limits are stronger for $\nu_e$ than for $\nu_\mu$ and in particular 
$\nu_\tau$, but are still rather weak, requiring $m_{\nu N}$ to be about an 
order of magnitude below $m_N$ for the relevant mass range~\cite{Bolton:2019pcu, Bryman:2019bjg,
deGouvea:2015euy, Kobach:2014hea, Abada:2013aba}. Indeed, saturating the
laboratory limits on the mixing angle is possible without terrible tuning of
$m_{\nu N}$. However, keeping the mixing angle below cosmological limits while
retaining a large neutrino magnetic moment again requires tuning of $m_{\nu
N}$.

\textbf{A leptoquark model} --- As a particular class of UV-complete theory
featuring large $\mu_\nu$ at one loop, let us consider models with TeV-scale
leptoquark (LQs). We assume the dominant LQ coupling is with the third family
of quarks, such that $\mu_{\nu}$ does not suffer a large Yukawa suppression
from the quark running in the loop. Another independent motivation to consider
this setup comes from flavor physics. In particular, third generation
leptoquarks are prime candidates for addressing the ongoing anomalies in
$B$-meson decays~\cite{Lees:2013uzd, Hirose:2016wfn, Aaij:2015yra,
Aaij:2014ora, Aaij:2017vbb, Aaij:2013qta, Aaij:2015oid, Aaij:2019wad,
Buttazzo:2017ixm}.  A scalar leptoquark $S_1$ with $SU(3)_c \times SU(2)_L
\times U(1)_Y$ quantum number $(\bar {\bf 3}, {\bf 1}, 1/3)$ is a prominent
successful model proposed in ref.~\cite{Bauer:2015knc}. The relevant Lagrangian
is
\begin{align}
  \mathcal{L}_{S_1}
    \supset y_1 \, \overline{b_R^{\ c}} N_R \, S_1
          + y_2 \, \overline{Q_{L}^3} L^{i\; c}_L \, S^\dagger_1 + \text{h.c.}\,.
  \label{eq:L-lq-1}
\end{align}
where $Q^3_L$, $L^i_L$, $b_R$, and $N_R$ are a left-handed SM quark doublet of
the third generation, a left-handed lepton doublet  of flavor $i$, a
right-handed bottom quark, and a right-handed neutrino, respectively, while the
superscript $c$ denotes a charge-conjugated field. Here, $y_1$ and $y_2$ are
dimensionless Yukawa couplings. It is implied that the $SU(2)$ (anti)doublets
$\psi$ and $\chi$ are contracted with the two-dimensional Levi-Civita tensor,
$\psi\, \chi \equiv \psi_{1} \chi_{2} - \psi_{2} \chi_{1}$. The leading
one-loop contribution to the $\nu_i$--$N_R$ transition neutrino magnetic moment is
\begin{align}
  \mu_{\nu} \approx \frac{e \, y_1 y_2}{8 \pi^2 m_{LQ}^2} \, m_{b} \log{\frac{m_b^2}{ m_{LQ}^2} } \,,
  \label{eq:magmom}
\end{align}
where $m_b$ is the bottom mass, and $m_{LQ}$ is the mass of the LQ. Thus, for
$y_{1} y_{2} \approx 0.05$ and $m_{LQ} \approx 1$~TeV, one obtains $|\mu_{\nu}|
\approx \SI{3e-11}{\mu_B}$, which is indeed within reach of XENON1T. Larger
$\mu_{\nu}$ would require larger Yukawa couplings; given the existing collider
searches for leptoquarks at the LHC (discussed for instance in the recent
review~\cite{Dorsner:2016wpm}), choosing $m_{LQ}$ below a TeV is not an option.

In any case, as already pointed out, a sizeable contribution to the Dirac mass
term $m_{\nu N}$ is generated by the diagrams in \cref{fig:diagram} (with the photon
line removed) and has to be tuned against the tree-level neutrino mass term.

\textbf{Muon $g-2$} --- Before we address this problem, let us first make a
comparison with the charged-lepton magnetic moments, focusing on the muon case.
The leptoquark interaction in eq.~\eqref{eq:L-lq-1} will induce a contribution
to the anomalous muon magnetic moment $\Delta a_\mu$ suppressed by the muon
mass. For $m_{LQ} \sim \SI{1}{TeV}$ and $y_{2} \sim 1$, the contribution to
$\Delta a_\mu$ is about an order of magnitude smaller than the observed
discrepancy between the theoretical prediction and the  measurement by the E821
experiment at BNL~\cite{Bennett:2006fi}. In other words, in this model,
neutrino transition magnetic moments observable in XENON1T are not in conflict
with measurements from the charged-lepton sector. In fact, the current anomaly
in muon $g-2$ can be fully accommodated if additional interactions are present.
The $S_1$ leptoquark has another independent gauge invariant interaction,
$\mathcal{L} \supset y'_1\,  \overline{t_R^{\ c}} e^i_R \, S_1 $, with the
right-handed top quark and a charged lepton. Even a small coupling, $y'_1 \sim
10^{-3}$ is enough to accommodated $\Delta a_\mu$, owing to the chiral
enhancement from the top quark in the loop~\cite{Dorsner:2019itg}. The
corresponding one-loop correction to the muon mass in this case is below the
tree-level one. In passing we note that an alternative 
muon $g-2$ explanation with leptoquarks was studied in \cite{ColuccioLeskow:2016dox} (see also recent Ref. \cite{Gherardi:2020qhc}).\\

\textbf{Voloshin mechanism with leptoquarks} --- We now review a mechanism for
breaking the relation \cref{eq:mu-mass-relation} between neutrino magnetic
moments and neutrino masses, and we implement this mechanism in the context of
TeV-scale leptoquarks.  The basic observation is that, due to the Lorentz
structure of the defining operators, the neutrino mass matrix is symmetric
while the magnetic moment matrix is antisymmetric in flavor space.  This
feature is exploited by the Voloshin mechanism for one-loop
models~\cite{Voloshin:1987qy}. Voloshin postulated an approximate global
$SU(2)_H$ symmetry, under which $(\nu_L^{\;c}, N_R)$ transforms as a doublet,
thus allowing the magnetic moment term $\bar{N}_R \sigma^{\mu \nu} \nu_L -
\overline{\nu_L^{\;c}} \sigma^{\mu \nu} N_R^{\;c}$, which is an $SU(2)_H$
singlet, while forbidding the $SU(2)_H$ triplet mass term $\bar{N}_R \nu_L +
\overline{\nu_L^{\;c}} N_R^{\;c}$. 

To realize this mechanism in the leptoquark scenario, we add in addition to
$S_1$ a second leptoquark $\tilde{R}_2\equiv ({\bf 3}, {\bf 2}, 1/6)$, with
interactions 
\begin{align}
  \mathcal{L}_{\tilde R_2}
    \supset - y_1 \, \tilde R^\dagger_2 \,\, \overline{b_R^{\;c}} L^{i c}_L
            + y_2 \, \overline{Q_L^3} N_R \, \tilde{R}_2 + \text{h.c.}\,.
  \label{eq:L-lq-2}
\end{align}
Note that $\mathcal{L} \supset \mathcal{L}_{\tilde R_2} + \mathcal{L}_{S_1}$
respects the global $SU(2)_H$ symmetry under which $(\tilde{R}_2^{-1/3},
S_1^\dagger)$ and $(\nu_L^{\;c}, N_R)$ transform as doublets. The $SU(2)_H$
symmetry of this sector is preserved by QCD and QED, but is explicitly broken
by $SU(2)_L \times U(1)_Y$ gauge interactions and by the lepton Yukawa
couplings.  In the limit of an exact $SU(2)_H$, the Dirac mass term $m_{\nu N}$
is exactly zero, while a transition magnetic moment is generated. In other
words, the two diagrams \cref{fig:diagram} (without the photon) cancel exactly since the
$SU(2)_H$ symmetry requires the leptoquarks to be mass-degenerate, and the
product of the couplings in the red vertices to be equal and opposite in the
two diagrams. For the neutrino magnetic moment (\cref{fig:diagram} with the
photon included), the relative sign between the two contributing diagrams is
flipped thanks to the opposite electric charges of the particles to which the
photon line is attached. Note that $SU(2)_H$ encompasses lepton number, thus
the leptoquark loops will not generate a Majorana mass term for neutrinos.

There are several terms in the Lagrangian that explicitly break $SU(2)_H$
symmetry:
\begin{enumerate}
  \item The charged lepton Yukawa couplings. This source of $SU(2)_H$ breaking
    is small thanks to the smallness of the charged lepton masses.

  \item Electroweak radiative corrections. These terms, related to the
    different $SU(2)_L \times U(1)_Y$ quantum numbers within the $SU(2)_H$
    multiplets, are somewhat more important.  Radiative corrections are still
    suppressed by $\alpha / (4 \pi)$, so while they do generate a neutrino
    Dirac mass $m_{\nu N}$, it is suppressed by more than three orders of
    magnitude compared to its value in leptoquark models without the Voloshin
    mechanism.  In other words, the cancellation of the diagrams in
    \cref{fig:diagram} is still spoiled, but only at the two-loop level.

  \item The Majorana mass term for the right-handed neutrino, which breaks not
    only $SU(2)_H$, but also lepton number, thus feeding into the active
    neutrino masses by the type-I seesaw formula.
\end{enumerate}
In spite of these $SU(2)_H$ breaking terms, even for an MeV-scale
right-handed neutrino only a mild cancellation between the loop-induced and
tree level mass terms is needed to obtain the correct active neutrino mass
scale and to sufficiently suppress the mixing angle.  If somewhat more
fine-tuning is accepted, it is also possible to generate active neutrino
masses predominantly through yet another $SU(2)_H$ breaking source, namely by
introducing an independent term $(\bar L_L \tilde{H}) (\tilde{H}^T
L_L^{\;c})$. Finally, if the inverse seesaw mechanism is invoked, neutrino
masses and mixing angles can be completely decoupled.

\textbf{$B$-meson anomalies} --- To connect the discussion to far to various
hints for lepton flavor universality violation in $B$-decays, note that these
hints can be explained by a scalar leptoquark with SM quantum numbers $(\bar
{\bf 3}, {\bf 1}, 1/3)$, coupled predominantly to third generation
quarks~\cite{Lees:2013uzd, Hirose:2016wfn, Aaij:2015yra, Aaij:2014ora,
Aaij:2017vbb, Aaij:2013qta, Aaij:2015oid, Aaij:2019wad, Buttazzo:2017ixm}.
This is precisely our $S_1$ leptoquark. In fact, the $S_1$ leptoquark can act
as a mediator both in neutral current transitions, $b \to s \ell^+ \ell^-$, and
in charged current transitions, $b \to c \tau  \nu$.  Here we discuss the two
cases separately. 

The charged current anomaly in
\begin{align}
  R(D^{(*)}) \equiv \frac{\BR(B \to D^{(*)} \tau\nu)}
                         {\BR(B \to D^{(*)} \mu\nu)}
\end{align}
requires leptoquark couplings to tau leptons ($i=3$ in \cref{eq:L-lq-1}),
implying a connection with the $\nu_\tau$ transition magnetic moment. The
effect on $R(D^{(*)})$ is generated by tree-level leptoquark exchange between
the quark and lepton currents.  Two different scenarios are possible depending
on the neutrino into which the $B$-meson decays. If in \cref{eq:L-lq-1} one
imposes $y_1 \ll y_2$, decays into $\nu_\tau$ will dominate, as suggested in
refs.~\cite{Bauer:2015knc, Cai:2017wry}.  The preferred parameter range is
roughly $m_{LQ} \sim \mathcal{O}(\si{TeV})$ and $y_2 \sim \mathcal{O}(1)$. In
the other scenario, the coupling $y_1$ in \cref{eq:L-lq-1} dominates, so the
$B$-meson decays into the sterile neutrino $N_R$, see
refs.~\cite{Robinson:2018gza, Azatov:2018kzb}. Interestingly enough, the
parameter range of interest for both, $B$-decays and the neutrino magnetic
moment, has a sizeable overlap. 

The neutral current anomaly in the ratio
\begin{equation}
  R(K^{(*)}) \equiv \frac{\BR(B \to K^{(*)}\mu^+\mu^-)}
                         {\BR(B \to K^{(*)} e^+e^-)}\Big|_{q^2_{{\rm min}} < q^2 < q^2_{{\rm max}}}~,
\end{equation}
requires a leptoquark coupling to muons or electrons ($i=2$ or $i=1$
in \cref{eq:L-lq-1}), implying a connection with $\nu_\mu$ or $\nu_e$
transition magnetic moment.  In
refs.~\cite{Bauer:2015knc, Becirevic:2016oho,Cai:2017wry,Angelescu:2018tyl}, the dominant effect in $b \to s
\mu^+\mu^-$ transitions comes from a box diagram at one-loop induced by the
$y_2$ coupling from \cref{eq:L-lq-1}. Interestingly, the anomalies are again
explained for $m_{LQ} \sim \mathcal{O}(\si{TeV})$ and $y_2 \sim
\mathcal{O}(1)$, the same parameter range required to generate a sizeable
neutrino magnetic moment. Note that the additional presence of the
$\tilde{R}_2$ leptoquark does not give corrections to $R(K^{(*)})$ if only the
interactions in \cref{eq:L-lq-2} are present. Another direction to be explored
is to utilize $\tilde{R}_2$ at tree-level, introducing a small leptoquark
coupling with the right-handed strange quark, as done for example in
ref.~\cite{Becirevic:2015asa}. A detailed analysis of flavor physics
constraints on the generation of neutrino magnetic moments is left for future work.

\textbf{Magnetic moment $\nu_e \to \nu_\mu$ with leptoquarks} --- 
Let us finally discuss how transition magnetic moments among the active
neutrinos can be generated in models without heavier sterile states.  In most
models, this requires severe tuning to avoid too large neutrino masses.  As
shown in ref.~\cite{Bell:2005kz,Chala:2020pbn}, operator mixing under renormalization sets a
stringent naturalness bound for active neutrinos of Dirac type. This bound is
avoided in the case of transition magnetic moments if the active neutrinos are
Majorana~\cite{Davidson:2005cs, Bell:2006wi}. One interesting model generating
a transition magnetic moment $\mu_{\nu_e \nu_\mu}$ between the $\nu_e$ and
$\nu_\mu$ flavors without undue tuning was recently proposed in
ref.~\cite{Babu:2020ivd}.  In this model, $SU(2)_H$ symmetry is not explicitly
broken by gauge interactions -- the colorless scalars $\eta^a \equiv ({\bf 1},
{\bf 1}, 1)$ and $\Phi^a \equiv ({\bf 1}, {\bf 2}, 1/2)$ come in two copies
($a=1,2$) as $SU(2)_H$ doublets, separately. The required $\eta$--$\Phi$ mass
mixing is introduced via the Higgs mechanism. This model therefore has less
$SU(2)_H$ breaking and thus more effectively protects the small neutrino
masses.  Since there are no $N_R$ states in ref.~\cite{Babu:2020ivd},
active--sterile mixing is not a concern either. Interestingly, the leading
$SU(2)_H$ breaking due to the muon Yukawa coupling elegantly generates the
correct neutrino mass scale.  We would like to point out that the same can be
achieved with our leptoquarks.  Let $S^a_1$ and $\tilde R_2^a$  ($a=1,2$) both
be doublets of $SU(2)_H$.  Their mass mixing is given by the lepton number
violating operator $H^\dagger \tilde R_2^a \epsilon^{a b} S_1^b$, while the
role of the $\tau$ lepton in ref.~\cite{Babu:2020ivd} is replaced by the bottom
quark.  More precisely, the $y_2$ interaction in \cref{eq:L-lq-1} and the $y_1$
interaction in \cref{eq:L-lq-2} are invariant under $SU(2)_H$ when $(L_e,
L_\mu)$ are combined into an $SU(2)_H$ doublet. This is an alternative to the
model of ref.~\cite{Babu:2020ivd} for simultaneously realizing small neutrino
masses and large $\mu_{\nu_e \nu_\mu}$, with quite different collider and
flavor phenomenology worth exploring.

To complete the discussion, we note that the leptoquark model for $\mu_{\nu_e \nu_\mu}$
can be embedded in the context of RPV SUSY, see refs.~\cite{Barbieri:1990qj,
Babu:1990wv} for an early work in this direction. Also, for recent studies
on the neutrino mass generation with $S_1$ and $\tilde{R}_2$ see
refs.~\cite{Helo:2015fba, Pas:2015hca, Hagedorn:2016dze, Dorsner:2017wwn}.

\section{Conclusions}
\label{sec:conclusions}

In summary, we have discussed neutrino magnetic moments in a broad context,
highlighting in particular the following take-home messages:
\begin{enumerate}
  \item While most of the existing literature on this topic is focused on magnetic
    moments coupling sub-eV states, transition magnetic moments with right-handed
    neutrinos at larger mass are equally well motivated, and offer a much richer
    phenomenology.

  \item Direct dark matter searches offer superb sensitivity to neutrino magnetic
    moments. Our analysis of XENON1T data sets some of the strongest limits
    on the parameter space spanned by the magnetic moment $\mu_\nu$ and the RH
    neutrino mass $M_N$.

  \item A transition magnetic moments of order $\mu_\nu \simeq \SI{6e-11}{\mu_B}$
    (for coupling to muon neutrinos only)
    with a RH neutrino mass $M_N \simeq \SI{100}{keV}$ might explain the XENON1T
    anomaly if conservative assumptions are adopted for the SN1987A and CMB
    $N_\text{eff}$ limits.

  \item Strong constraints on the parameter space are imposed by stellar cooling
    and BBN. We have carried out in particular a detailed study of the latter.
    Other cosmological constraints, which might spoil the explanation of the XENON1T
    anomaly, are less robust: the limit from SN1987A is avoided if the
    neutrino flux from this supernova was dominated by accretion onto the supernova
    core rather than cooling of the core itself~\cite{Bar:2019ifz}.
    The CMB measurement of $N_\text{eff}$ needs to be taken with a grain of salt
    in view of the persisting $H_0$ tension, which prefers $N_\text{eff} > 3$.

  \item Neutrino magnetic moments and neutrino masses, both manifestations of
    chirality-flipping interactions, are typically interlinked in a high-energy
    theory. Thus, large magnetic moments are difficult to reconcile with the
    observed smallness of neutrino masses.
    This conclusion can, however, be avoided, as we have illustrated in
    a TeV-scale leptoquark model that has been recently proposed as a solution
    to $B$-physics anomalies and the muon $g-2$ anomaly.  We have implemented
    the Voloshin mechanism~\cite{Voloshin:1987qy} in this model by postulating
    that the two leptoquarks appearing in it are members of a doublet under an
    approximate $SU(2)_H$ horizontal symmetry.  As a result, the phenomenology
    of the right-handed neutrinos is dominated by the transition magnetic moment
    rather than the active--sterile mixing, as commonly assumed in the literature.
\end{enumerate}
Most of our results are concisely summarized in \cref{fig:constraints},
which collects the various constraints on neutrino magnetic moments that
we have discussed.

We believe that because of their rich phenomenology and manifold connections
to other areas of particle physics, astrophysics, and cosmology, neutrino
transition magnetic moments are a prime target for current and future
direct dark matter searches, given that these detectors will be able to set
the most stringent terrestrial limits.  Using these detectors in that way highlights
once more that running modern underground experiments does not mean sitting
next to a tank of liquid and waiting, as some critics have claimed in the past.
Rather, it means operating a multi-purpose observatory with a rich and diverse
physics program, and with results that will reverberate throughout many
domains of fundamental physics for years to come.

\section*{Acknowledgments}

First and foremost, we would like to thank Frederik Depta, Marco Hufnagel, and
Kai Schmidt-Hoberg, the authors of ref.~\cite{Depta:2020wmr} for sharing their
BBN code and for their invaluable assistance.  We are moreover indebted to
Roni Harnik and Pedro Machado for many insightful discussions on the XENON1T
anomaly, and to Simon Knapen, Matthew McCullough, Ennio Salvioni, Bibhushan
Shakya, Marko Simonovi\'{c} for useful comments.  JK's work has been partially
supported by the European Research Council (ERC) under the European Union's
Horizon 2020 research and innovation program (grant agreement No.\ 637506,
``$\nu$Directions''). The work of AG is partially supported by the European
Research Council (ERC) under the European Union’s Horizon 2020 research and
innovation programme, grant agreement 833280 (FLAY). 


\bibliographystyle{JHEP}
\bibliography{refs}

\end{document}